\documentclass[aps,prl,superscriptaddress,showpacs,showkeys,reprint]{revtex4-1}
\usepackage{amsmath}
\usepackage{amsfonts}
\usepackage{amsthm}
\usepackage{graphics}
\usepackage{graphicx}
\usepackage{subfigure}
\usepackage{amssymb}
\usepackage{color}
\usepackage{url}
\usepackage{hyperref}
\usepackage[abs]{overpic}

\begin{document}

\title{Homogeneous Isotropic Superfluid Turbulence in Two Dimensions:
Inverse and Forward Cascades in the Hall-Vinen-Bekharevich-Khalatnikov
model}

\author{Vishwanath Shukla}
\email{research.vishwanath@gmail.com}
\affiliation{Centre for Condensed Matter Theory, Department of Physics,
Indian Institute of Science, Bangalore 560012, India}
\author{Anupam Gupta} 
\email{anupam1509@gmail.com }
\affiliation{Department of Physics, University of Rome “Tor Vergata”,
Via della Ricerca Scientifica 1, 00133 Rome, Italy}
\author{Rahul Pandit}
\email{rahul@physics.iisc.ernet.in}
\altaffiliation[\\ also at~]{Jawaharlal Nehru Centre For Advanced
Scientific Research, Jakkur, Bangalore, India.}
\affiliation{Centre for Condensed Matter Theory, Department of Physics, 
Indian Institute of Science, Bangalore 560012, India.} 

\date{\today}
\begin{abstract}
We present the first direct-numerical-simulation study of the statistical properties
of two-dimensional superfluid turbulence in the
Hall-Vinen-Bekharevich-Khalatnikov two-fluid model. We show that both
normal-fluid and superfluid  energy spectra can exhibit two power-law regimes,
the first associated with an inverse cascade of energy and the second with the
forward cascade of enstrophy. We quantify the mutual-friction-induced alignment
of normal and superfluid velocities by obtaining probability distribution
functions of the angle between them and the ratio of their moduli.
Our study leads to specific suggestions for experiments.
\end{abstract}
\pacs{47.27.Gs, 47.27.ek, 47.37.+q, 67.25.dm}
\keywords{superfluid; turbulence; two-fluid model}
\maketitle

The elucidation of superfluid turbulence, a problem of central importance in quantum
fluids and nonequilibrium statistical mechanics, continues to provide challenges for
experiments, theory, and numerical
simulations~\cite{donnellybook,lathrop2011review,skrbeksreeni2012review,marcpnas14}.
Such turbulence has been
studied more often in three dimensions (3D) than in two dimensions (2D).
It is well known that 2D and 3D \textit{fluid} turbulence are
qualitatively
different~\cite{frischbook,lesieurbook,boffetta2012review,pramanareview};
similar differences have not been explored in detail for
\textit{superfluid} turbulence. Therefore, we initiate a study of the statistical
properties of 2D homogeneous, isotropic, superfluid turbulence, at the level
the Hall-Vinen-Bekharevich-Khalatnikov (HVBK), two-fluid
model~\cite{donnelly1999cryogenicreview,barenghi1983mfriction,marcpnas14,hall1956rotation,khalatnikov},
with the specific goal of elucidating the natures of both inverse and forward
cascades of energy and enstropy, the mean square vorticity. Homogeneous, isotropic, 2D and 3D
fluid turbulence are essentially different because, in the former, both
the energy and the enstrophy are conserved in the inviscid, unforced
limit, whereas, in the latter, only the energy is
conserved~\cite{frischbook,lesieurbook,boffetta2012review,pramanareview}.
Therefore, in 2D fluid turbulence, energy, injected at a wave number
$k_{\rm f}$, shows an inverse cascade towards large length scales 
(wave number $ k < k_{\rm f}$), whereas the enstrophy
displays a forward cascade to small
length scales ($ k > k_{\rm f}$); these inverse and forward cascades yield,
respectively, energy spectra that scale as $E(k) \sim k^{-5/3}$ and $E(k)
\sim k^{-\delta}$, where $\delta$ depends on the friction ($\delta = 3$
if there is no friction). By contrast, 3D fluid turbulence shows only a
forward cascade of energy with $E(k) \sim k^{-5/3}$, at the level of
Kolmogorov's (K41) phenomenological theory~\cite{frischbook} and for
$k_f \ll k \ll k_d$, where $k_d$ is the wave number scale at which
viscous dissipation becomes significant.

Our direct numerical simulation (DNS), which we have designed to study the
statistical properties of inverse and forward cascades in the HVBK model,
yields several interesting results that have not been anticipated hitherto: (1)
Both normal-fluid and superfluid energy spectra, $E^n(k)$ and $E^s(k)$,
respectively, show inverse- and forward-cascade power-law regimes.
(2) The forward-cascade power
law depends on (a) the friction coefficient, as in 2D fluid turbulence, and, in
addition, on (b) the coefficient $B$ of mutual friction, which couples normal
and superfluid velocities.  (3) As $B$ increases, the normal and superfluid
velocities, ${\bf u}_n$ and ${\bf u}_s$, respectively, tend to get locked to each
other, and, therefore, $E^s(k) \simeq E^n(k)$, especially in the
inverse-cascade regime.  (4) We quantify this locking tendency by calculating
the probability distribution functions (PDFs) $P(\cos(\theta))$ and
$P(\gamma)$, where the angle $\theta \equiv \cos^{-1}((\mathbf{u}_{\rm n}\cdot
\mathbf{u}_{\rm s})/(|\mathbf{u}_{\rm n}||\mathbf{u}_{\rm s}|))$ and the
amplitude ratio $\gamma=|\mathbf{u}_{\rm n}|/|\mathbf{u}_{\rm s}|$; the former
has a peak at $\cos(\theta) = 1$; and the latter exhibits a peak at $\gamma =
1$ and power-law tails on both sides of this peak. (5) This locking increases
as we increase $B$, but the power-law exponents for the tails of
$P(\gamma)$ are universal, in so far as they do not depend on $B$,
$\rho_n/\rho$, where $\rho_n$ and $\rho$ are normal-fluid and total densities,
respectively, and $k_{\rm f}$. 

The incompressible, 2D HVBK equations are
~\cite{donnelly1999cryogenicreview,barenghi1983mfriction,marcpnas14,hall1956rotation,khalatnikov}
\begin{subequations}
\label{eq:2d2fluidu}
\begin{align}
D_t\mathbf{u}_{\rm n}
& = -\frac{1}{\rho_{\rm n}}\nabla p_{\rm n} + \nu_{\rm n}\nabla^2\mathbf{u}_{\rm n}
 - \mu_{\rm n}\mathbf{u}_{\rm n} 
+ \mathbf{F}^{\rm n}_{\rm mf} + \mathbf{f}^{\rm n}_u, \label{eq:2d2fluidu_n}\\
D_t\mathbf{u}_{\rm s}
& = -\frac{1}{\rho_{\rm s}}\nabla p_{\rm s} + \nu_{\rm s}\nabla^2\mathbf{u}_{\rm s}
 - \mu_{\rm s}\mathbf{u}_{\rm s} 
+ \mathbf{F}^{\rm s}_{\rm mf} + \mathbf{f}^{\rm s}_u, \label{eq:2d2fluidu_s}
\end{align}
\end{subequations}
where $D_t\mathbf{u}_i\equiv\partial_t+\mathbf{u}_i\cdot\nabla$,
$\nabla\cdot\mathbf{u}_i=0$ is the incompressibility condition, and the
subscript $i \in ({\rm n,s})$ denotes the normal fluid ({\rm n}) or the
superfluid ({\rm s}); $\rho_i$, $p_i$, and $\nu_i$ are the density,
partial pressure, and viscosity, respectively, of the component $i$. 
Linear-friction terms, with coefficients $\mu_i$, model
air-drag-induced or bottom friction. For the superfluid
$\nu_{\rm s}$ and $\mu_{\rm s}$ are zero, but any DNS
study must use $\nu_{\rm s} (\neq0) \ll \nu_{\rm n}$ and 
$\mu_{\rm s}\ll \mu_{\rm n}$ to avoid numerical instabilities and to achieve a
statistically steady state. The mutual-friction terms can be
written as $\mathbf{F}^{\rm n}_{\rm mf} = (\rho_s/\rho)\mathbf{f}_{\rm
mf}$ and $\mathbf{F}^{\rm s}_{\rm mf} = -(\rho_n/\rho)\mathbf{f}_{\rm
mf}$ in Eqs.~(\ref{eq:2d2fluidu_n}) and (\ref{eq:2d2fluidu_s}),
respectively, where
\begin{equation}\label{eq:mf}
\mathbf{f}_{\rm mf} = \frac{B}{2}\frac{\mathbf{\omega}_{\rm s}}{|\mathbf{\omega}_{\rm s}|}\times
(\mathbf{\omega}_{\rm s}\times\mathbf{u}_{\rm ns})
+ \frac{B'}{2}\mathbf{\omega}_{\rm s}\times\mathbf{u}_{\rm ns},
\end{equation} 
with $\mathbf{u}_{\rm ns} = (\mathbf{u}_{\rm n} - \mathbf{u}_{\rm s})$ the slip
velocity, and $B$ and $B'$ the coefficients of mutual friction.  In most of our
studies we set $B'=0$ so, in 2D, $\mathbf{f}_{\rm mf} =
-\frac{B}{2}|\omega_{\rm s}|\mathbf{u}_{\rm ns}$. (We have checked in one
representative case that our results do not change qualitatively if $B' > 0$.)
In our DNS, we use the stream-function $\psi_i$ and vorticity $\omega_i=
\nabla\times\mathbf{u}_i = -\nabla^2\psi_i$
formulation~\cite{perlekarnjp2dfilms}. To obtain a statistically steady state,
we force the vorticity field with a Kolmogorov-type term
$f^i_{\omega}=-f^i_0k^i_{\rm f}\cos(k^i_{\rm f}x)$, where $f^i_0$ and
$k^i_{\rm f}$ are the amplitude and the forcing wave number, respectively.
We use (a) $k_{\rm f}^i = 2$ and (b) $k_{\rm f}^i = 50$;
the former leads to energy spectra that are dominated by a forward cascade of
enstrophy, whereas the latter yields spectra with an inverse cascade of energy
and a forward cascade of enstrophy; we force the dominant component in case (b)
(i.e., the normal-fluid (superfluid) component if $\rho_n/\rho > 0.5$
($\rho_n/\rho \leq 0.5$)).

We perform a DNS of Eqs.~(\ref{eq:2d2fluidu_n}) and (\ref{eq:2d2fluidu_s}) with
periodic boundary conditions, on a square simulation domain with side $L=2\pi$,
by using a pseudospectral method~\cite{fornberg1998,perlekarnjp2dfilms} 
with $N_c^2$ collocation points and the $2/3$
dealiasing rule. For time evolution we use a second-order, exponential time
differencing Runge-Kutta method~\cite{cox2002etd}.  The
parameters of our DNS runs are given in 
Table~\ref{table:para} (and in the Supplemental Material~\cite{supplement}). 
We characterize our system by computing the spectra
$E_n(k)$ and $E_s(k)$, $E_i(k) = \langle\sum_{k-\frac{1}{2}<k' \leq k+\frac{1}{2}}
\lvert\mathbf{u}_{\rm i} (\mathbf{k'},t)\rvert^2\rangle_t$
($\langle\rangle_t$ denotes a time average), the probability distribution
functions (PDFs) $P(\omega_i)$ of the vorticities and $P(\cos(\theta))$, the
cumulative PDF $Q(\gamma)$ of $\gamma$, energy and enstrophy fluxes $\Pi_{\rm
i}(k,t)$ and $Z_{\rm i}(k)$ ($i \in ({\rm n, s})$), respectively,
and the mutual-friction transfer function $M_{\rm i}(k)$, which we
define below.
\begin{table}
\begin{center}
\small
\resizebox{\linewidth}{!}{
   \begin{tabular}{@{\extracolsep{\fill}} c c c c c c c c c c c c c c c c c c c c}
    \hline

    $ $ & $\rho_n/\rho$ & $B$ & $\nu_n$ & $\nu_s$ & 
$k^{\rm n}_f$ & $k^{\rm s}_f$ & $f^{\rm n}_0$ & $f^{\rm s}_0$ & 
$Re^n_{\lambda}$ & $Re^s_{\lambda}$ \\
   \hline \hline

{\tt R0}  & $-$ & $-$ & $10^{-4}$ & $10^{-5}$ & 
$2$ & $2$ & $10^{-3}$ & $10^{-3}$  & 
$92.77$ & $1.25\times10^{3}$ \\

{\tt R1}  & $0.1$ & $1.0$ & $10^{-4}$ & $10^{-5}$ & 
$2$ & $2$ & $10^{-3}$ & $10^{-3}$  & 
$112.9$ & $1.3\times10^{3}$ \\

{\tt R2a} & $0.1$ & $1.0$ & $10^{-4}$ & $10^{-5}$ & 
$-$ & $50$ & $-$ & $10^{-1}$ & 
$108.4$ & $876.7$ \\

{\tt R2b} & $0.1$ & $2.0$ & $10^{-4}$ & $10^{-5}$ & 
$-$ & $50$ & $-$ & $10^{-1}$ & 
$100.6$ & $876.8$ \\

{\tt R2c} & $0.1$ & $5.0$ & $10^{-4}$ & $10^{-5}$ & 
$-$ & $50$ & $-$ & $10^{-1}$ & 
$94.3$ & $876.5$\\

{\tt R3 } & $0.05$ & $1.0$ & $10^{-4}$ & $10^{-5}$ & 
$-$ & $50$ & $-$ & $10^{-1}$ & 
$119.1$ & $976.7$ \\

{\tt R4 } & $0.3$ & $1.0$ & $10^{-4}$ & $10^{-5}$ & 
$-$ & $50$ & $-$ & $10^{-1}$ & 
$62.9$ & $487.6$ \\

{\tt R5 } & $0.5$ & $1.0$ & $10^{-5}$ & $10^{-6}$ & 
$-$ & $50$ & $-$ & $10^{-1}$ & 
$484.1$ & $4.1\times10^{3}$ \\

{\tt R6 } & $0.9$ & $1.0$ & $10^{-5}$ & $10^{-6}$ & 
$50$ & $-$ & $10^{-1}$ & $-$ & 
$617.0$ & $7.19\times10^{3}$ \\

\hline
\end{tabular}
}
\end{center}
\caption{\small Parameters for our DNS runs $\tt R0$-$\tt R6$ with 
$N_c^2=1024^2$ collocation points:
$\rho_n/\rho$ is the fraction of the normal fluid, $B$ the
mutual-friction coefficient, $\nu_n$ ($\nu_s$) the normal-fluid (superfluid) kinematic viscosity,
and $k^n_{\rm f}$ ($k^s_{\rm f}$) and $f^n_0$ ($f^s_0$)
are the forcing wave vector and the forcing amplitude for the normal fluid
(superfluid).
The coefficient of linear friction for the normal fluid (superfluid)
$\mu_n=10^{-2}$ ($\mu_s=5\times10^{-3}$) is kept fixed. For more parameters see Table~1 in 
~\cite{supplement}.
} 
\label{table:para}
\end{table}

\begin{figure}[floatfix]
\centering
\resizebox{\linewidth}{!}{
\includegraphics[width=0.49\linewidth]{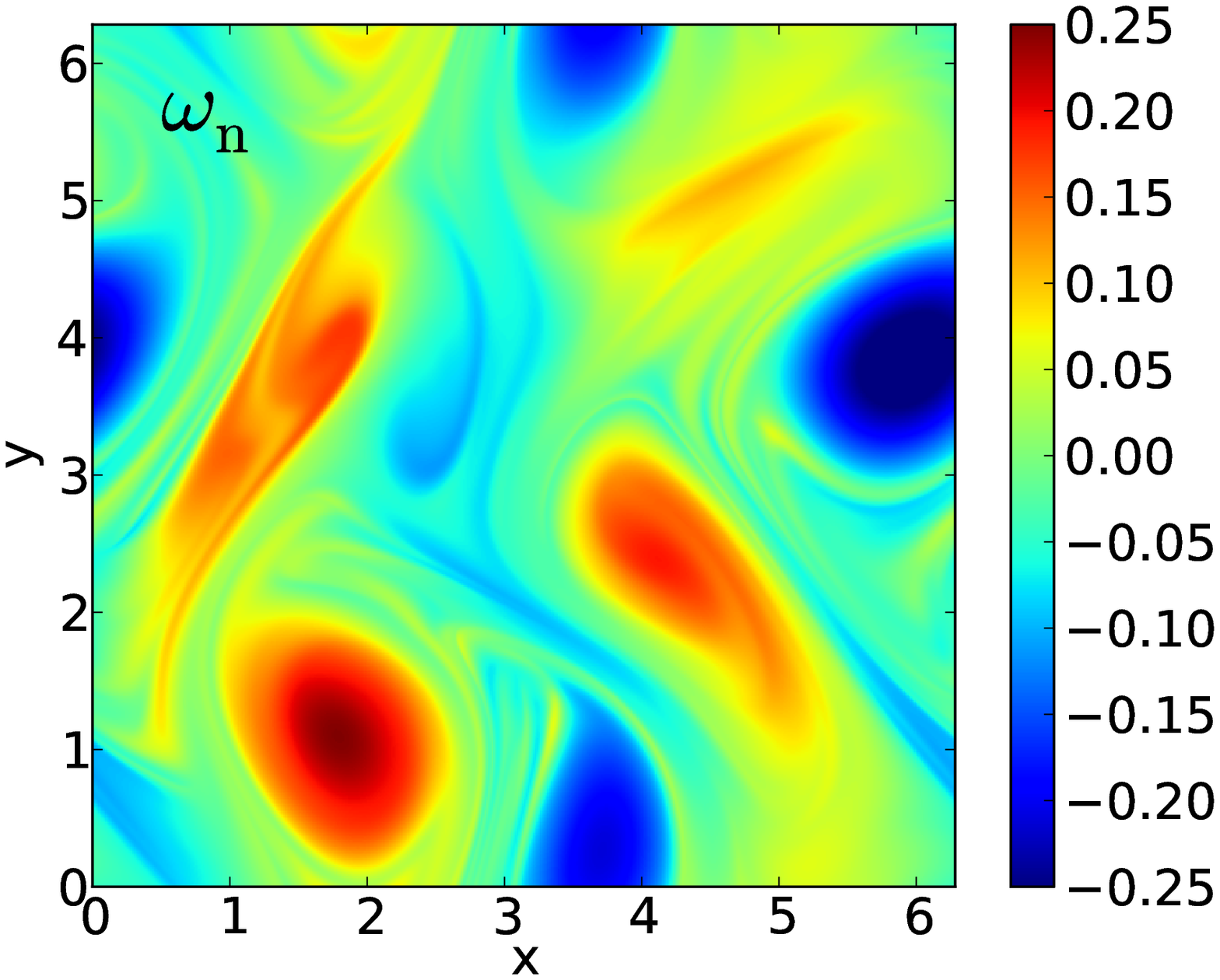}
\put(-50,75){{\bf (a)}}
\includegraphics[width=0.49\linewidth]{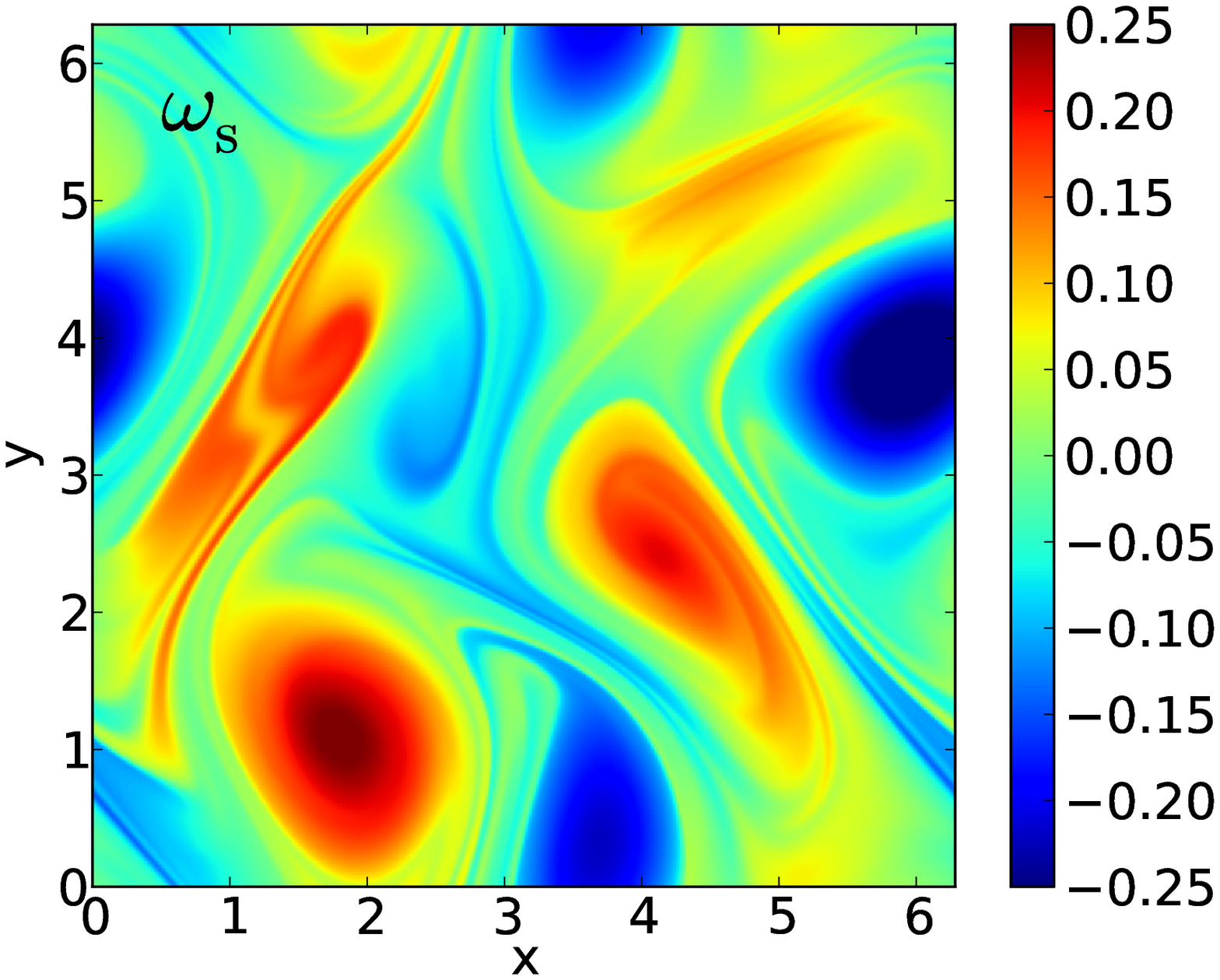}
\put(-50,75){{\bf (b)}}
}
\resizebox{\linewidth}{!}{
\includegraphics[width=0.4\linewidth,unit=1mm]{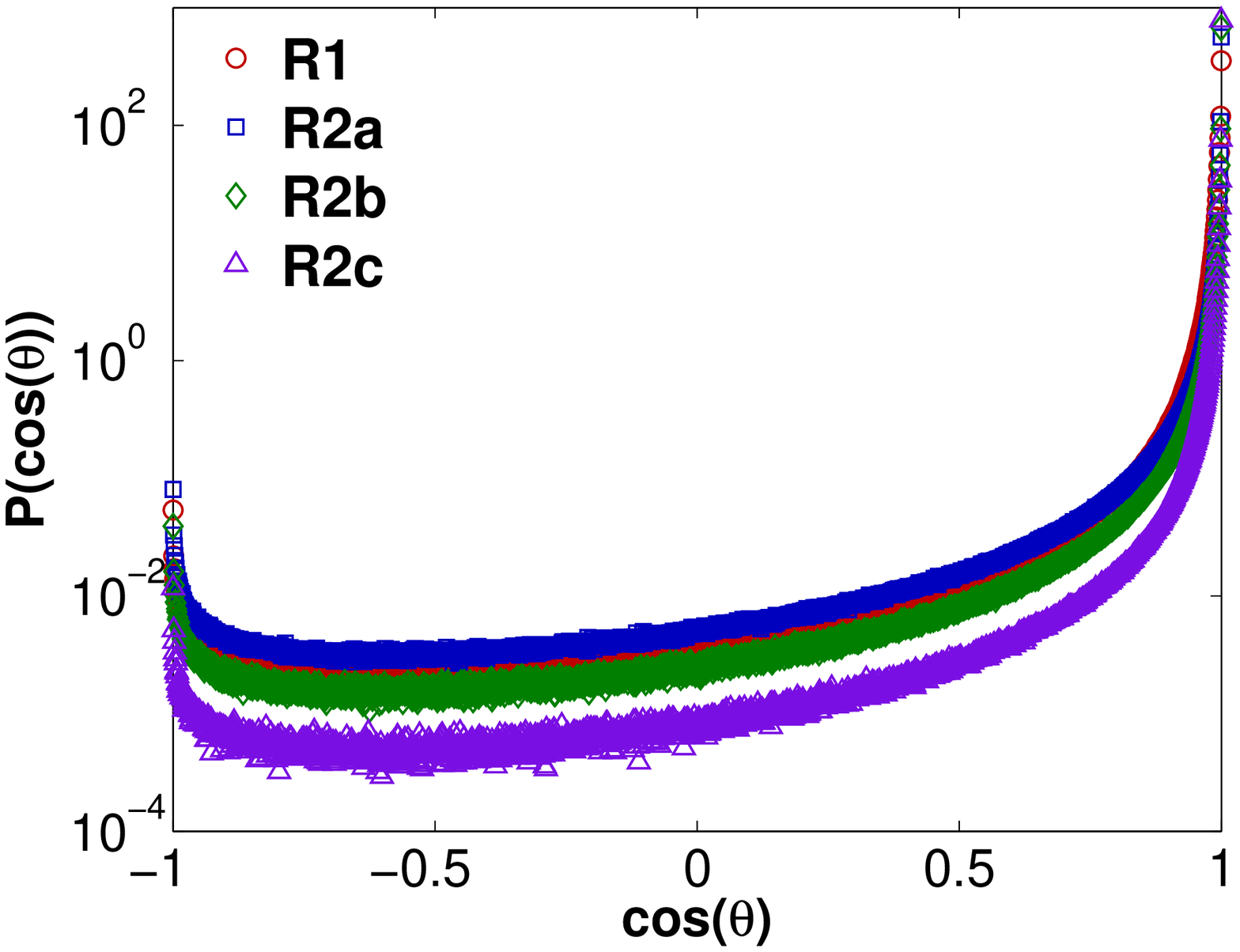}
\put(-30,55){{\bf (c)}}
\includegraphics[width=0.4\linewidth,unit=1mm]{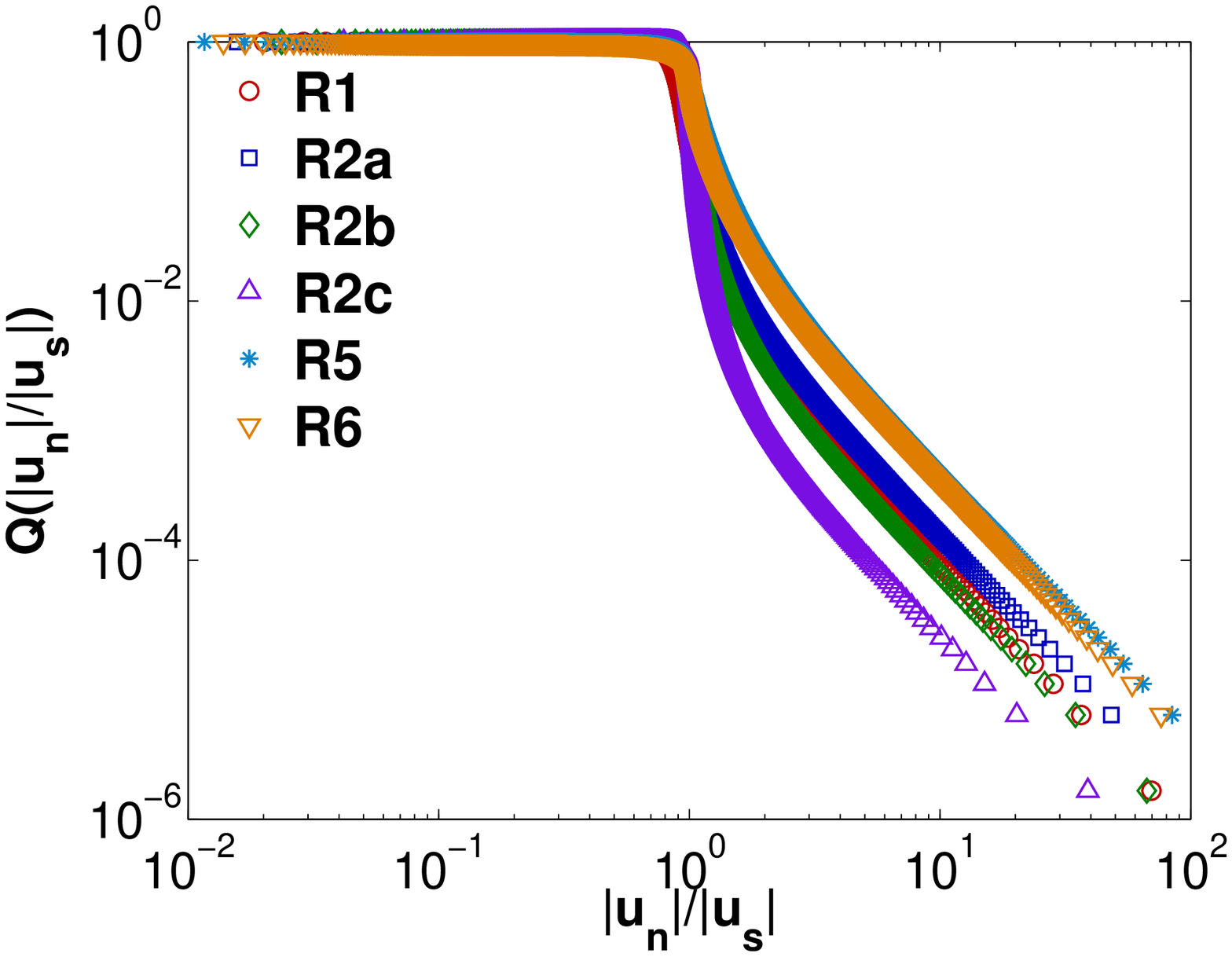}
\put(-30,55){{\bf (d)}}
}
\caption{(Color online) Pseudocolor plots of the vorticity fields, $\omega_n$
and $\omega_s$, from our DNS run $\tt R1$ at $t=1720$ (panels (a) and (b),
$k_{\rm f}=2$); these plots show that the normal and superfluid
component are locked to each other.
(c) Semilogarithmic (base 10) plots of the PDF
$P(\cos(\theta))$ of the angle $\theta$ between $\mathbf{u}_{\rm n}$ and
$\mathbf{u}_{\rm s}$ for runs $\tt R1$ (red circles), $\tt R2a$ ($B=1$, blue
squares), $\tt R2b$ ($B=2$, green diamonds), and $\tt R2c$ ($B=5$, purple
triangles).
(d) Log-log (base 10) plots of the complementary cumulative distribution functions (CDF)
$R(\gamma)$ of $\gamma=|\mathbf{u}_{\rm n}|/|\mathbf{u}_{\rm s}|$ for the runs $\tt R1$,
$\tt R2a-R2c$, $\tt R5$, and $\tt R6$. These CDFs show power-law tails 
($R(\gamma) \sim \gamma^{-2}$) that imply $P(\gamma) \sim \gamma^{-3}$, for $\gamma \gg 1$.
}
\label{fig:snapvort_pdf}
\end{figure}

In Fig.~\ref{fig:snapvort_pdf} we present pseudocolor plots of
$\omega_n$ and $\omega_s$ for run $\tt R1$ (panels (a) and (b)).
Similar plots for run $\tt R2a$ with $k_{\rm f} = 50$ are given in Fig.~1 in the 
Supplemental Material~\cite{supplement}. The sizes of the
vortical regions in these plots are $\sim k_{\rm f}^{-1}$ (as in 2D fluid
turbulence with friction~\cite{boffetta2012review,perlekarnjp2dfilms}).
Figures~\ref{fig:snapvort_pdf} (a) and (b) show that the normal
and superfluid components are nearly locked to each other; this is illustrated
dramatically in Video M1
~\cite{supplement}, in which the lower two panels show the spatiotemporal
evolution of Figs.~\ref{fig:snapvort_pdf} (a) and (b) and the top
two panels show how $\omega_n$ and $\omega_s$ evolve in the absence of mutual
friction (i.e., $B=B'=0$); in the latter case, $\omega_n$ evolves to a frozen,
stationary state;
however, if $B > 0$, then the turbulence in the superfluid component is
transferred to the normal component (top two panels of Video M1). Such a
transfer of turbulence has been envisaged in 3D superfluid
turbulence~\cite{roche2009HVBK3d,morris2008vortexlock,wacks2011STshellmodel}
but never displayed as graphically as in our Video M1.

We quantify the locking of the normal and superfluid velocities by plotting, in
Fig.~\ref{fig:snapvort_pdf}~(c), for the illustrative runs $\tt R1$ and
$\tt R2a$-$\tt R2c$, the PDF $P(\cos(\theta))$, which shows
a peak at $\cos(\theta)=1$ and falls rapidly with increasing $\theta$; this
indicates that $\mathbf{u}_{\rm n}(\mathbf{r},t)$ and $\mathbf{u}_{\rm
s}(\mathbf{r},t)$ align preferentially along the same direction; the degree of
alignment increases as we increase $B$. 
In Figs.~\ref{fig:snapvort_pdf} (d) we show, respectively,
plots of the complementary cumulative distribution functions (CDFs) $R(\gamma)$ of
$\gamma=|\mathbf{u}_{\rm n}|/|\mathbf{u}_{\rm s}|$, 
for the runs $\tt R1$,  $\tt R2a$-$\tt R2c$, $\tt
R5$, and $\tt R6$. These CDFs exhibit power-law tails that imply that 
$P(\gamma) \sim \gamma^{-3}$, for $\gamma \gg 1$ (A similar analysis of the left tail
~\cite{supplement} yields $P(\gamma) \sim \gamma$, for $\gamma \ll 1$).
The power-law exponents of these tails of $P(\gamma)$ are
universal in the sense that they do not depend on $B$, $\rho_{\rm n}/\rho$, and
$k_{\rm f}$.

\begin{figure*}
\centering
\begin{overpic}
[width=0.3\linewidth,unit=1mm]{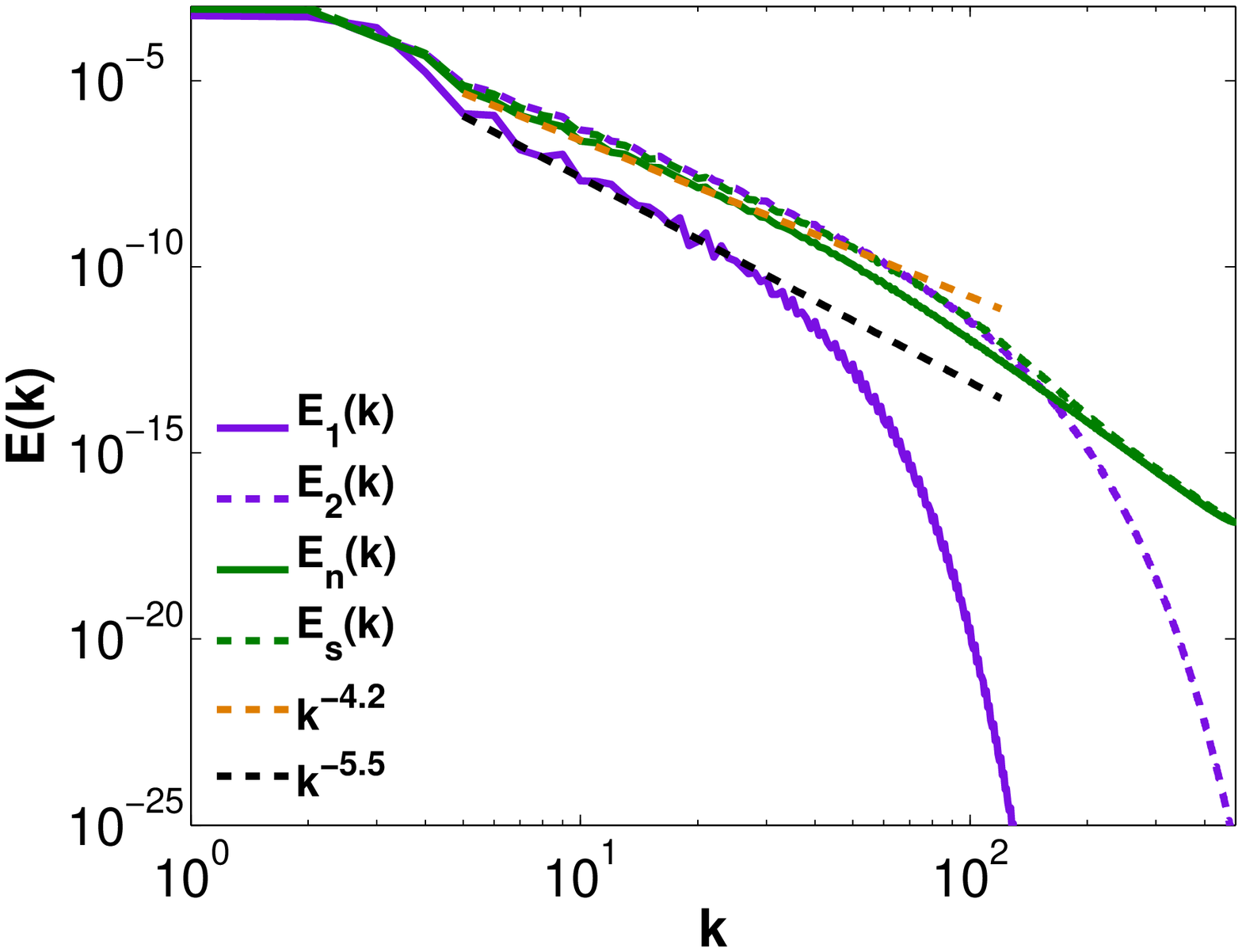}
\put(25,10){\large{\bf (a)}}
\end{overpic}
\begin{overpic}
[width=0.3\linewidth,unit=1mm]{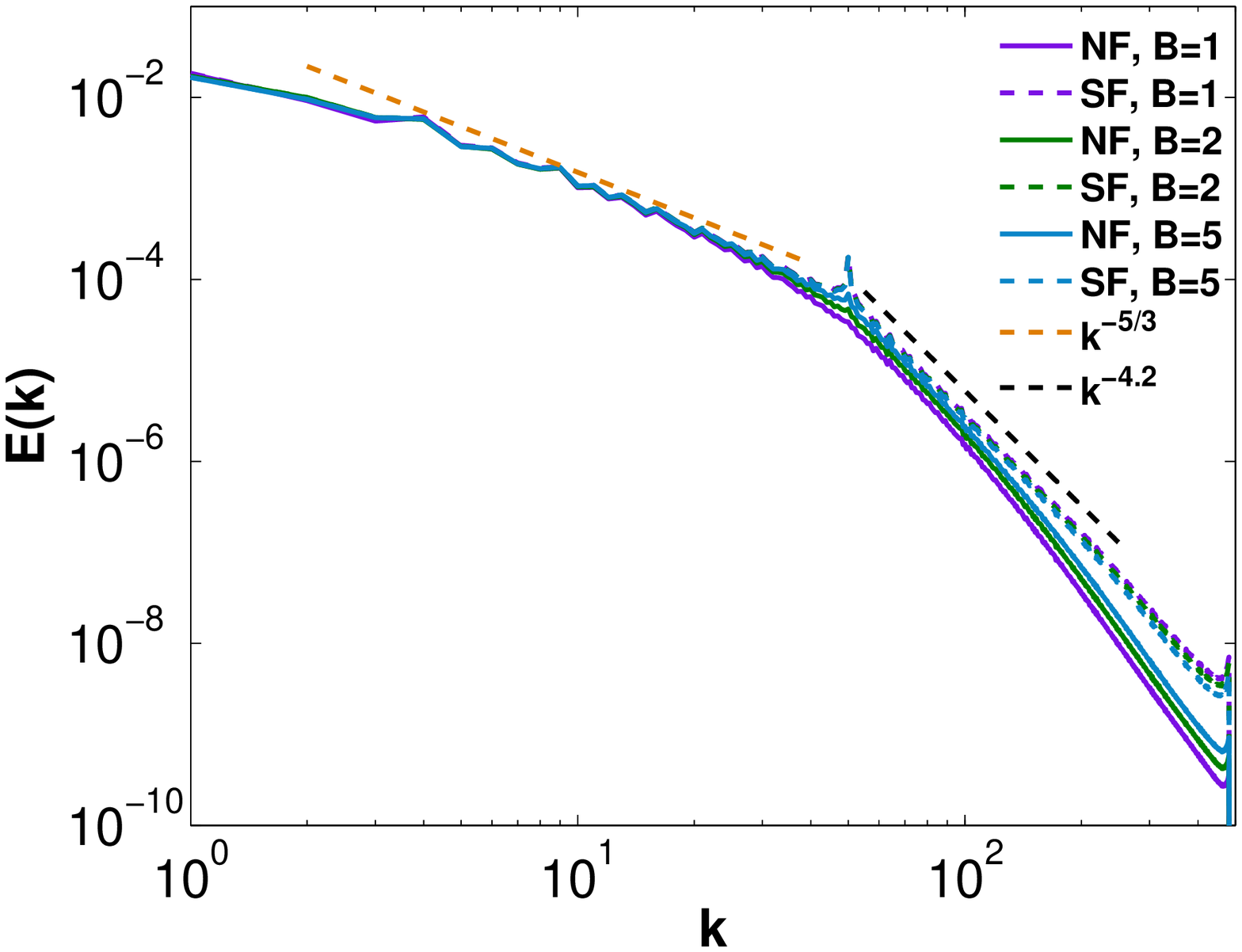}
\put(25,10){\large{\bf (b)}}
\end{overpic}
\begin{overpic}
[width=0.3\linewidth,unit=1mm]{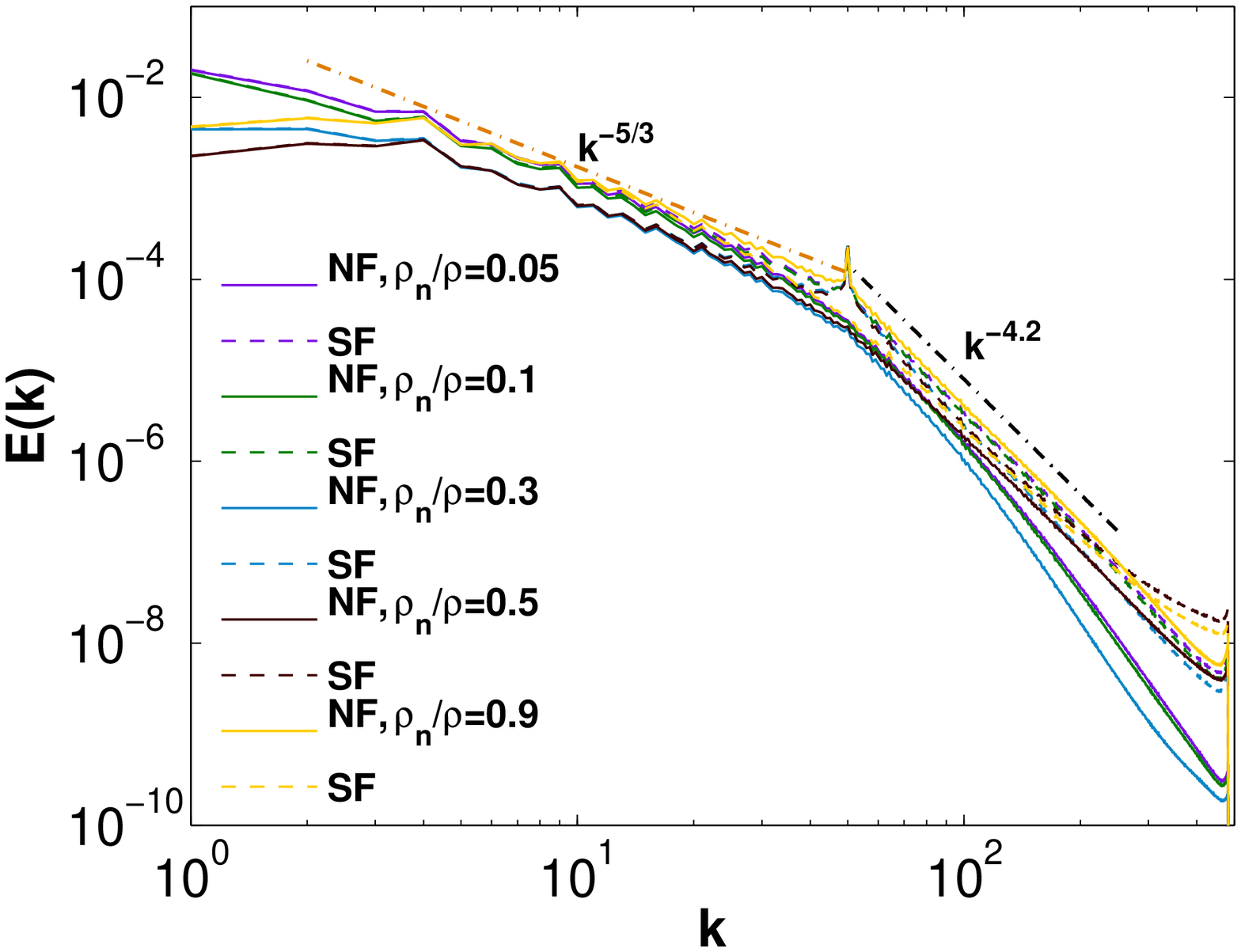}
\put(25,10){\large{\bf (c)}}
\end{overpic}
\\
\begin{overpic}
[width=0.3\linewidth,unit=1mm]{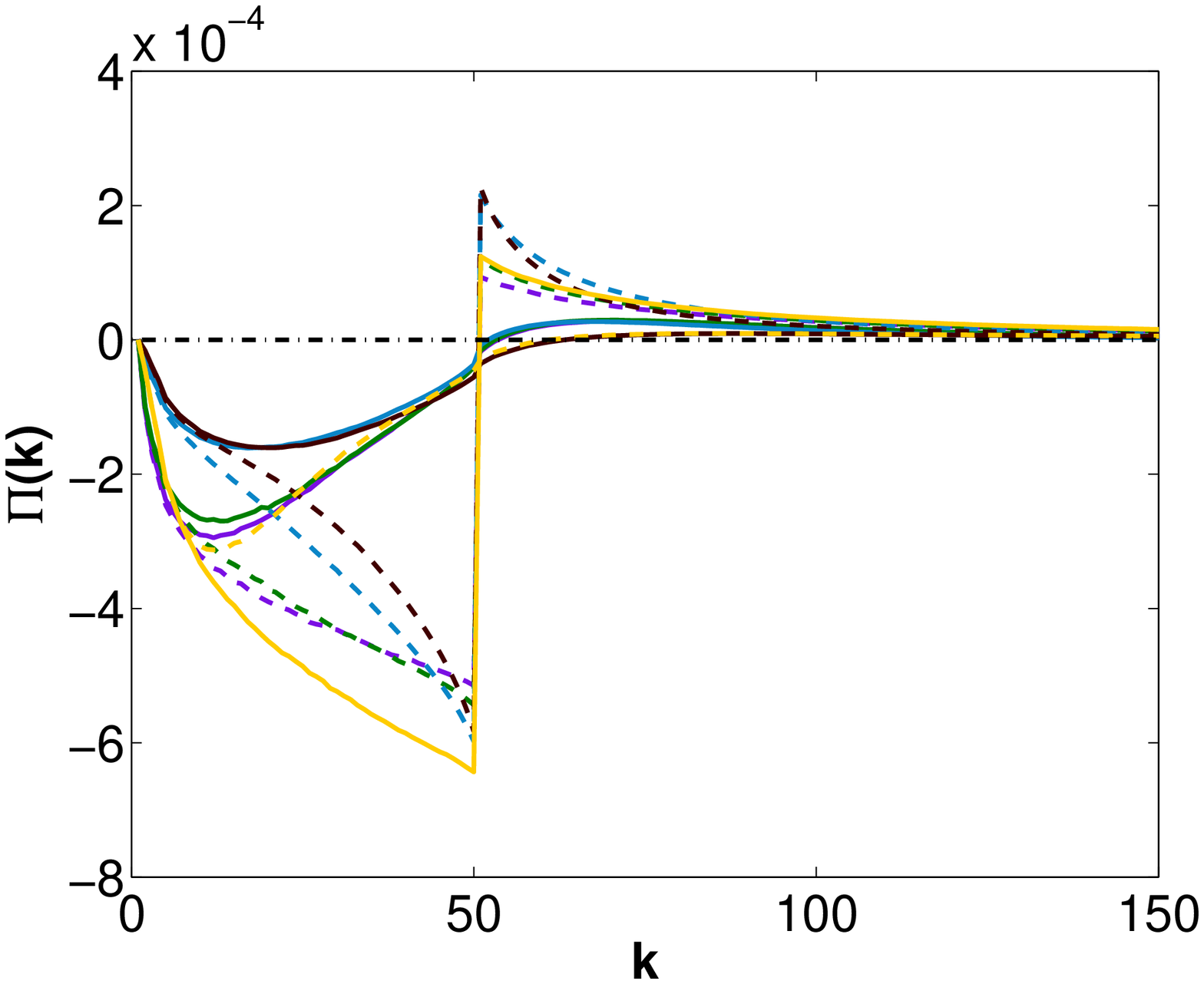}
\put(25,10){\large{\bf (d)}}
\end{overpic}
\begin{overpic}
[width=0.3\linewidth,unit=1mm]{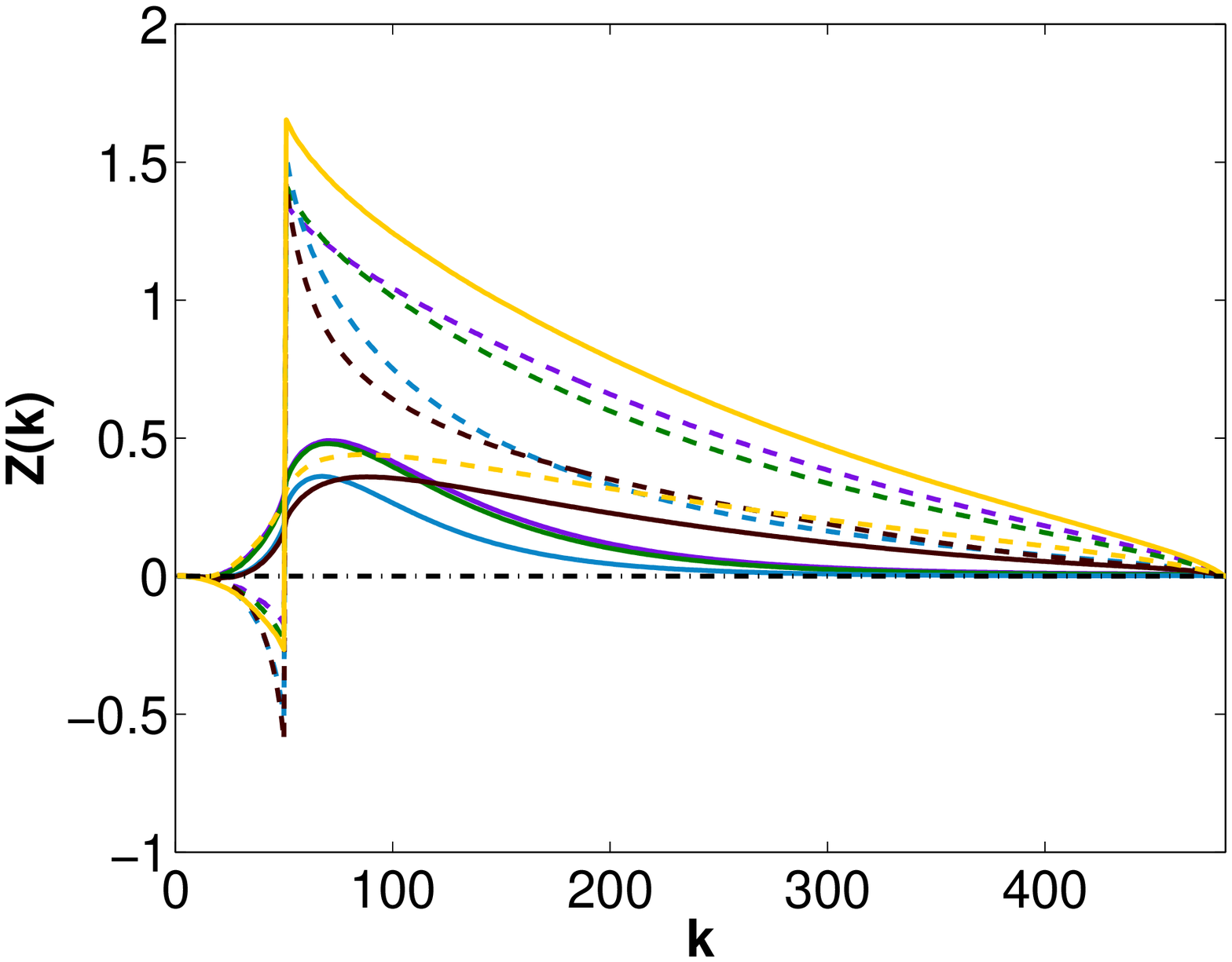}
\put(25,10){\large{\bf (e)}}
\end{overpic}
\begin{overpic}
[width=0.3\linewidth,unit=1mm]{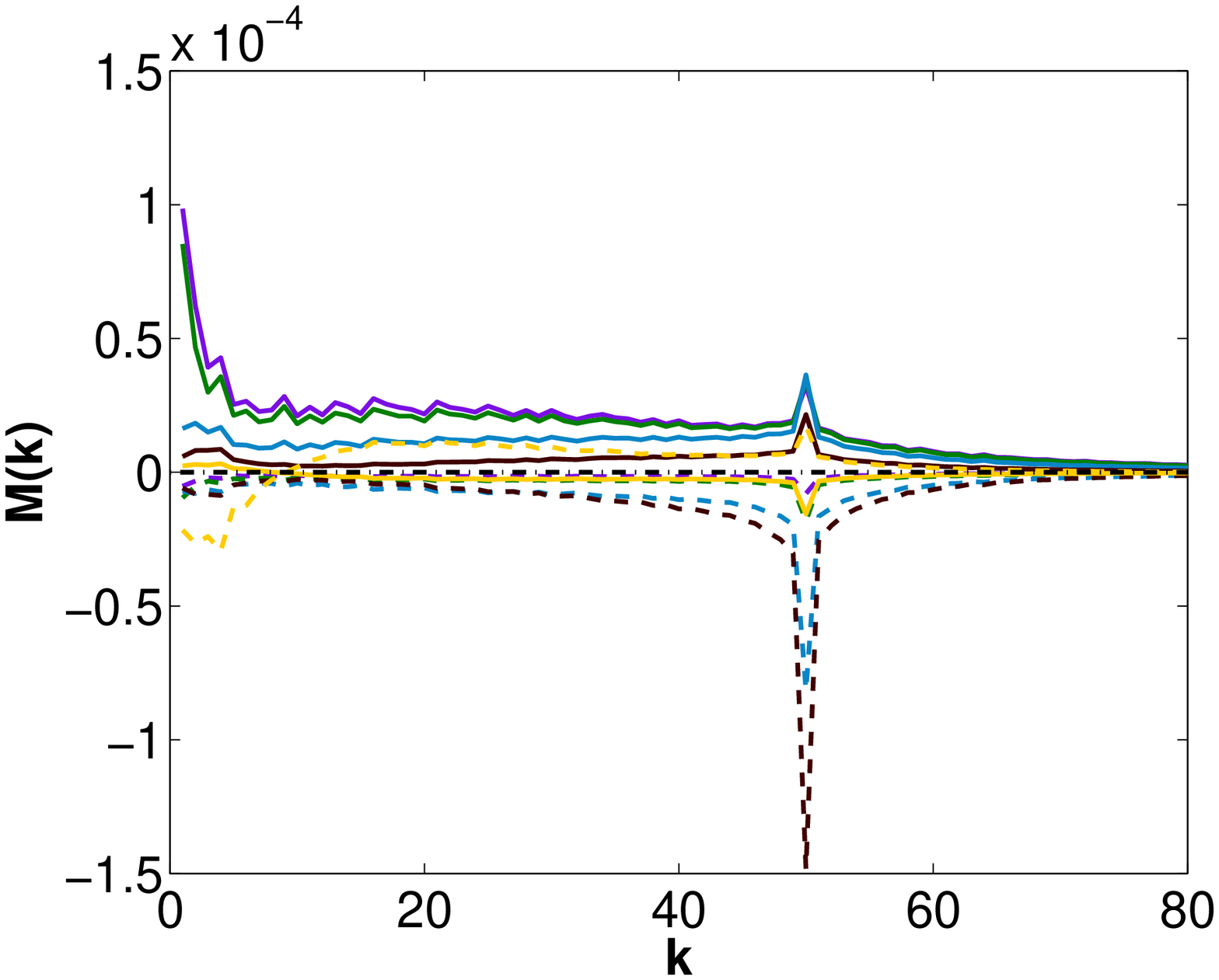}
\put(25,10){\large{\bf (f)}}
\end{overpic}
\caption{(Color online) [Top panels] Log-log plots of the energy spectra
$E_n(k)$ (full lines) and $E_s(k)$ (dashed lines) from our DNS runs: (a) $\tt
R0$ ($B=0$, purple lines) and $\tt R1$ ($B=1$, green lines) with $k_{\rm
f}=2$; (b) $\tt R2a$ ($B=1$, purple curves), $\tt R2b$ ($B=2$, green curves),
and  $\tt R2c$ ($B=5$, sky-blue curves) with $k^s_{\rm f}=50$ and
$\rho_n/\rho=0.1$; (c) $\tt R2a$ (purple curves), $\tt R3$ (green curves), $\tt
R4$ (sky-blue curves), $\tt R5$ (black curves), and $\tt R6$ (yellow curves),
with $B=1$; we force the dominant component.  [Lower panels] Plots of (d)
the energy flux $\Pi_i(k)$, (e) the enstrophy flux $Z_i(k)$, and (f) the
mutual-friction transfer function $M_i(k)$, for the DNS runs
represented in (c), with the same color codes as mentioned above. 
The abbreviation NF (SF) stands for normal-fluid (superfluid).}
\label{fig:Espectra_flux}
\end{figure*}

Figure~\ref{fig:Espectra_flux} (a) compares energy spectra from runs $\tt
R0$ and $\tt R1$, in which  
there are no inverse-cascade regimes in energy spectra; this figure
illustrates how the mutual-friction-induced interaction between the two
components modifies the energy spectra $E_i(k,t)$.
For the run $\tt R0$, in which $B=0$ and, therefore, the normal and
superfluid components are uncoupled, $E^n(k)$ and $E^s(k)$ are shown in
Fig.~\ref{fig:Espectra_flux} (a) by full and dashed purple lines,
respectively: the
power-law regimes, more prominent in $E^s(k)$ than in $E^n(k)$, are
characterized by different, apparent scaling exponents, because the
normal-fluid Reynolds number is too small for fully developed,
normal-fluid turbulence. When we couple the normal and superfluid
components, as in the run $\tt R1$, $E^n(k)$ (green full curve in
Fig.~\ref{fig:Espectra_flux} (a)) is pulled up towards $E^s(k)$ (green
dashed curve in Fig.~\ref{fig:Espectra_flux} (a)), by virtue of the
locking tendency that we have mentioned above; furthermore, both $E^n(k)$
and $E^s(k)$ now (i) display $k^{-\delta}$ forward-cascade, scaling
ranges, with $\delta \simeq 4.2$, (ii) lie very close to each other at
small wave numbers, and (iii) show dissipation regions at much higher wave
numbers than in their counterparts when there is no coupling ($B=0$ and
run $\tt R0$). 

To study dual cascades, i.e., (i) an inverse cascade of energy for $k<k_{\rm
f}$ and (ii) a direct cascade of enstrophy for $k>k_{\rm f}$, we use our
DNS runs $\tt R2a$-$\tt R6$ (see Table~\ref{table:para}).
Figure~\ref{fig:Espectra_flux} (b) shows $E^n(k)$ (full curves) and $E^s(k)$
(dashed curves) with dual cascades, for the runs $\tt R2a$ with $B=1$ (purple
curves), $\tt R2b$ with $B=2$ (green curves), and $\tt R2c$ with $B=5$ (blue
curves). The inverse-cascade inertial ranges (with $k<k_{\rm f}$) of $E^n(k)$
and $E^s(k)$ exhibit scaling that is consistent with a $k^{-5/3}$ form (orange,
dashed line), whereas the forward-cascade ranges (with $k>k_{\rm f}$) are
consistent with $k^{-\delta}$ scaling, and $\delta \simeq 4.2$ (black, dashed
line). In the forward-cascade regime of 2D fluid turbulence, the value of
$\delta$ depends on the coefficient of linear
friction~\cite{boffetta2012review,perlekarnjp2dfilms,ott,musacchio2005friction2d};
we find that, in the 2D HVBK model, $\delta$ depends not only on the
coefficients of linear friction, but also on $B$.
Furthermore, the locking that we have discussed above makes $E^n(k)$  and
$E^s(k)$ lie more-or-less on top of each other for a considerable range of wave
numbers; not surprisingly, this range of overlap increases as $B$ increases;
for $B=5$ it extends into the direct-cascade region.
Figure~\ref{fig:Espectra_flux} (c) shows inverse- and forward-cascade regimes
in log-log plots of $E^n(k)$ (full curves) and $E^s(k)$ (dashed curves) versus
$k$ for five representative values of $\rho_{\rm n}/\rho$ (runs R2a (purple
curves), R3 (green curves), R4 (blue curves), R5 (black curves), and R6 (yellow
curves)), with $B=1$ and $k<k_{\rm f} = 50$.

The HVBK model allows us to study the evolution of two-fluid turbulence as we
change $\rho_{\rm n}/\rho$, which is small at  low temperatures and increases
as the temperature increases and approaches the superfluid transition
temperature; if $\rho_{\rm n}/\rho=0.05$, 
HVBK turbulence is close to that of a pure superfluid, on the length and
Mach-number scales at which the HVBK model is valid; in contrast, HVBK
turbulence at $\rho_{\rm n}/\rho=0.9$ is close to that of a classical,
incompressible fluid. In Fig.~\ref{fig:Espectra_flux} (c), the orange,
dot-dashed line indicates a $k^{-5/3}$ power-law form that is visually close to
the slopes (in log-log plots) of the energy spectra in the inverse-cascade
scaling ranges; the black, dot-dashed line indicates a $k^{-4.2}$ power-law
form that is visually close to the slope of the $E^n(k)$ spectrum in the
forward-cascade scaling range for $\rho_{\rm n}/\rho=0.9$.
A complete study of the dependence of $\delta$ on $\mu_{\rm i}$ and $B$ 
requires extensive, and high-resolution DNS
studies whose current computational cost is prohibitive.

To characterize fluxes in the inverse- and forward-cascade regimes we use
the energy-transfer relations for 2D, homogeneous, isotropic, HVBK,
turbulence, namely, 
\begin{equation}\label{eq:transfer}
\partial_{\rm t} E_{\rm i}(k,t) = -\mathcal{D}_{\rm i}(k,t) + \mathcal{T}_{\rm i}(k,t) 
+ \mathcal{M}_{\rm i}(k,t) + \mathcal{F}^{\rm i}_o(k),
\end{equation}
where $i\in(n,s)$, $\mathcal{D}_{\rm i}(k,t) \equiv \sum_{k-\frac{1}{2}<k'\leq
k+\frac{1}{2}} (\nu_{\rm i} k'^2 + \mu_{\rm i})\lvert\mathbf{u}_{\rm
i}(\mathbf{k'})\rvert^2$ is the transfer function, which combines the effects
of viscous dissipation and the friction, $\mathcal{T}_{\rm i}(k,t)$ is the
kinetic-energy transfer because of the triad interactions of the Fourier
components of the velocities, and $\mathcal{F}^{\rm i}_o(k)$ is the
energy-injection spectrum for the component $i\in(n,s)$.
The mutual-friction-induced exchange of energy between the
normal and the superfluid components is measured by
\begin{equation}
\mathcal{M}_{\rm i}(k,t) \equiv \displaystyle\sum_{k-\frac{1}{2}<k'\leq k+\frac{1}{2}}
\mathbf{F}^{\rm i}_{\rm mf}(\mathbf{k'},t)\cdot\mathbf{u}_{\rm i}(-\mathbf{k'},t).
\end{equation}

The kinetic-energy fluxes, through the wave number $k$, are 
$\Pi_{\rm i}(k,t)=\langle\int^{\rm k_{max}}_k\mathcal{T}_{\rm i}(k',t)dk'\rangle_t$;
and their analogs for the enstrophy fluxes are  $Z_{\rm i}(k)$ ($i \in (n,s)$).
We plot these versus $k$ in
Figs.~\ref{fig:Espectra_flux} (d) and (e), respectively, for the same runs
($\tt R2a$ and $\tt R3$-$\tt R6$) and the same color codes as in
Fig.~\ref{fig:Espectra_flux} (c). In Fig.~\ref{fig:Espectra_flux} (d), for each
one of these runs, the energy fluxes $\Pi_{\rm i}(k) < 0$, for
$k<k_{\rm f}$, which confirms that we have inverse cascades of energy; in
contrast, the enstrophy fluxes $Z_{\rm i}(k) > 0$, for $k>k_{\rm
f}$, in Fig.~\ref{fig:Espectra_flux} (e), so we have  forward cascades of
enstrophy.  For the runs $\tt R2a$ and $\tt R3$-$\tt R6$ we plot, in
Fig.~\ref{fig:Espectra_flux} (f), the transfer functions 
$M_i(k)=\langle\mathcal{M}_{\rm i}(k,t)\rangle_t$ versus $k$,
which characterizes the energy exchange between the normal and superfluid components.

Other PDFs, e.g., those of velocity components and the vorticity, in 2D HVBK turbulence
are qualitatively similar to their classical-fluid-turbulence
counterparts~\cite{supplement,BarenghivpdfcellavgPRE}. We expect, as in the case of 3D superfluid
turbulence~\cite{roche2007vortdensityspectra,salort2010turbulent,salort2011investigation,
salort2012energy}, that the results from our 2D HVBK studies will be borne out
by experiments on 2D superfluid turbulence if these experiments probe length
scales that are larger than the mean separation between quantum vortices.  To
obtain power-law tails in velocity-component PDFs, of the type that have been seen in some
experiments in 3D quantum turbulence~\cite{paoletti2008vpdf}, we must use
either (a) the Gross-Pitaevskii (GP) equation
~\cite{white2010vpdf,nowak2012nonthermal,vmrnjp13}, which can resolve quantum
vortices, or (b) Biot-Savart-type models~\cite{adachi2011bsvpdf}.
Two-dimensional superfluid turbulence is now being studied numerically with
such models~\cite{tsubota20102dgpdirect,vmrnjp13,marcpnas14,reeves2013inverse}.
In particular, some DNS studies have looked for inverse cascades in 2D GP
turbulence, which is forced and in which a dissipation term is used to obtain a
statistically steady state. One such study~\cite{reeves2013inverse} has
obtained an inverse cascade.
On scales that are much larger than the mean separation between quantum
vortices, and when quantum vortices of the same sign cluster, we expect
superfluids to be described by the HVBK equations, if we restrict ourselves to
low-Mach-number
flows~\cite{barenghi1983mfriction,donnelly1999cryogenicreview,marcpnas14}; and
the extraction of HVBK-model parameters from GP studies is just
beginning to be studied in
3D~\cite{BerloffPRL2007disspdyna,giorgio2011PRBcounterflow,*girogio2011longPRE}
and 2D~\cite{Jackson2009PRAFiniteTvortex,vsthesis}.

Our DNS study of homogeneous, isotropic turbulence in the 2D HVBK model has led 
to the first elucidation of inverse and forward cascades in this system,
has contrasted them with their counterparts in 2D fluid turbulence, and led
to qualitatively new results that await experimental confirmation in turbulent
superfluid films. We have shown
that both  $E^n(k)$ and $E^s(k)$ exhibit inverse- and forward-cascade
power-law regimes. We have demonstrated that, as
$B$ increases, ${\bf u}_n$ and ${\bf u}_s$ tend to align with each other:
the PDF $P(\cos(\theta))$ has a peak at $\cos(\theta) = 1$ and
$P(\gamma)$ displays power-law tails with universal exponents,
which do not depend on $B$, $\rho_n/\rho$, and $k_{\rm f}$. 
The parameters $B$ and $\rho_n/\rho$ depend on
the temperature; and this dependence has been measured in 
experiments~\cite{donbarenghiexpt3dvalues98} in 3D; such experimental
studies have not been carried out in 2D.

We thank M.E. Brachet and A. Bhatnagar for discussions, CSIR, DST, and UGC (India) for 
financial support, and SERC (IISc) for computational resources.

\bibliographystyle{apsrev4-1}
\bibliography{references}

\clearpage

{\large\bf{Supplemental Material}}
\vspace{0.5cm}

In this Supplemental Material we give details of our calculations; these augment the results
that we have presented in the main part of this paper.

\vspace{0.5cm}
{\bf Video M1} (\url{http://youtu.be/-ZDkoxQInXY})

This video illustrates the spatiotemporal evolution, via pseudocolor plots, of
$\omega_n$ (left panels) and $\omega_s$ (right panels) in which the mutual friction is 
(a) absent in the top two panels (DNS run $\tt R0$) and (b) present in the lower two 
panels (DNS run $\tt R1$).

In Table~\ref{table:paraSM} we give the detailed list of parameters, which we use in our DNS
runs.  
The energy and enstrophy are defined as 
$E_{\rm i}=\frac{1}{2}\sum_kE_{\rm i}(k)$ and $\Omega_{\rm i}=\frac{1}{2}\sum_kk^2E_{\rm
i}(k)$ ($i \in ({\rm n,s})$), respectively. The root-mean-square velocity is 
$u^{\rm i}_{\rm rms}=\sqrt{E_{\rm i}}$; the Taylor microscale is
\begin{equation}
\ell^{\rm i}_{\lambda}=\sqrt{\frac{E_{\rm i}}{2\Omega_{\rm i}}};
\end{equation}
the Taylor-microscale Reynolds number is
\begin{equation}
Re^{\rm i}_{\lambda}=\frac{u^{\rm i}_{\rm rms}\ell^{\rm i}_{\lambda}}{\nu_{\rm i}};
\end{equation}
the integral length scale is 
\begin{equation}
l^{\rm i}_0 = \frac{\sum_kE_{\rm i}(k)/k}{E_{\rm i}};
\end{equation}
the eddy-turnover time is
\begin{equation}
\tau^{\rm i}_{\rm eddy}=\frac{l^{\rm i}_0}{u^{\rm i}_{\rm rms}};
\end{equation}
the dissipation scale is
\begin{equation}
\eta_{\rm i} = \Bigl[\frac{\nu^2_{\rm i}}{2\Omega_{\rm i}}\Bigr]^{1/4};
\end{equation}
here $i \in ({\rm n,s})$.

\begin{table*}
\begin{center}
\small
\resizebox{\linewidth}{!}{
   \begin{tabular}{@{\extracolsep{\fill}} c c c c c c c c c c c c c c c c c c c c}
    \hline

    $ $ & $N_c$ & $\rho_n/\rho$ & $B$ & $\nu_n$ & $\nu_s$ & $\mu_n$ & $\mu_s$ &
$k^{\rm n}_f$ & $k^{\rm s}_f$ & $f^{\rm n}_0$ & $f^{\rm s}_0$ & 
$\ell^{\rm n}_{\rm \lambda}$ & $\ell^{\rm s}_{\rm \lambda}$ & $Re^n_{\lambda}$ & $Re^s_{\lambda}$ 
& $\tau^{\rm n}_{\rm eddy}$ & $\tau^{\rm s}_{\rm eddy}$ & 
$k_{\rm max}\eta_{\rm n}$ & $k_{\rm max}\eta_{\rm s}$\\
   \hline \hline

{\tt R0}  & $1024$ & $-$ & $-$ & $10^{-4}$ & $10^{-5}$ & $10^{-2}$ & $5\times10^{-3}$ &
$2$ & $2$ & $10^{-3}$ & $10^{-3}$  & 
$0.36$ & $0.38$ & $92.77$ & $1.25\times10^{3}$ &
$51.1$ & $45.8$ & $17.7$ & $5.16$\\

{\tt R1}  & $1024$ & $0.1$ & $1.0$ & $10^{-4}$ & $10^{-5}$ & $10^{-2}$ & $5\times10^{-3}$ &
$2$ & $2$ & $10^{-3}$ & $10^{-3}$  & 
$0.371$ & $0.378$ & $112.9$ & $1.3\times10^{3}$ &
$46.3$ & $42.3$ & $16.8$ & $5.01$\\

{\tt R2a} & $1024$ & $0.1$ & $1.0$ & $10^{-4}$ & $10^{-5}$ & $10^{-2}$ & $5\times10^{-3}$ &
$-$ & $50$ & $-$ & $10^{-1}$ & 
$0.062$ & $0.049$ & $108.4$ & $876.7$ &
$5.43$ & $5.12$ & $2.89$ & $0.82$\\

{\tt R2b} & $1024$ & $0.1$ & $2.0$ & $10^{-4}$ & $10^{-5}$ & $10^{-2}$ & $5\times10^{-3}$ &
$-$ & $50$ & $-$ & $10^{-1}$ & 
$0.058$ & $0.05$ & $100.6$ & $876.8$ &
$5.23$ & $5.05$ & $2.80$ & $0.82$\\

{\tt R2c} & $1024$ & $0.1$ & $5.0$ & $10^{-4}$ & $10^{-5}$ & $10^{-2}$ & $5\times10^{-3}$ &
$-$ & $50$ & $-$ & $10^{-1}$ & 
$0.054$ & $0.05$ & $94.3$ & $876.5$ &
$5.12$ & $5.04$ & $2.7$ & $0.82$\\

{\tt R3 } & $1024$ & $0.05$ & $1.0$ & $10^{-4}$ & $10^{-5}$ & $10^{-2}$ & $5\times10^{-3}$ &
$-$ & $50$ & $-$ & $10^{-1}$ & 
$0.064$ & $0.051$ & $119.1$ & $976.7$ &
$5.03$ & $4.77$ & $2.84$ & $0.771$\\

{\tt R4 } & $1024$ & $0.3$ & $1.0$ & $10^{-4}$ & $10^{-5}$ & $10^{-2}$ & $5\times10^{-3}$ &
$-$ & $50$ & $-$ & $10^{-1}$ & 
$0.052$ & $0.039$ & $62.9$ & $487.6$ &
$5.72$ & $5.14$ & $3.18$ & $0.868$\\

{\tt R5 } & $1024$ & $0.5$ & $1.0$ & $10^{-5}$ & $10^{-6}$ & $10^{-2}$ & $5\times10^{-3}$ &
$-$ & $50$ & $-$ & $10^{-1}$ & 
$0.043$ & $0.035$ & $484.1$ & $4.1\times10^{3}$ &
$4.88$ & $4.49$ & $0.916$ & $0.264$\\

{\tt R6 } & $1024$ & $0.9$ & $1.0$ & $10^{-5}$ & $10^{-6}$ & $10^{-2}$ & $5\times10^{-3}$ &
$50$ & $-$ & $10^{-1}$ & $-$ & 
$0.039$ & $0.047$ & $617.0$ & $7.19\times10^{3}$ &
$3.50$ & $3.84$ & $0.771$ & $0.268$\\

\hline
\end{tabular}
}
\end{center}
\caption{\small Parameters for our DNS runs $\tt R0$-$\tt R6$:
$\rho_n/\rho$ is the fraction of the normal fluid, $B$ the
mutual-friction coefficient, $N_c^2$ the number of collocation points,
$\nu_n$ ($\nu_s$) the normal-fluid (superfluid) kinematic viscosity,
$\mu_n$ ($\mu_s$) the coefficient of linear friction for the normal fluid
(superfluid), and $k^n_{\rm f}$ ($k^s_{\rm f}$) and $f^n_0$ ($f^s_0$)
are the forcing wavevector and the forcing amplitude for the normal fluid
(superfluid); $\nu_s$ and $\mu_s$ should vanish for a superfluid but they
are included here for numerical stability, with $\nu_s \ll \nu_n$ and
$\mu_s \ll \mu_n$; $\ell^n_{\lambda}$ and $\ell^s_{\lambda}$ are the normal-fluid and
superfluid Taylor microscales, $Re^n_{\lambda}$ and $Re^s_{\lambda}$ the associated
Reynolds numbers, $\tau^{\rm n}_{\rm eddy}$ and $\tau^{\rm s}_{\rm eddy}$ the eddy-turnover
times; $\eta_{\rm n}$ and $\eta_{\rm s}$ are the dissipation length scales;
$k_{\rm max}$ is the magnitude of the largest wave vector in our $2/3$-dealiased DNS.
In runs $\tt R2a$-$\tt R6$ we force the dominant
component (i.e., the normal-fluid (superfluid) component if $\rho_n/\rho
> 0.5$ ($\rho_n/\rho \leq 0.5$)).} 
\label{table:paraSM}
\end{table*}

\begin{figure*}[floatfix]
\centering
\resizebox{\linewidth}{!}{
\includegraphics[width=0.49\linewidth]{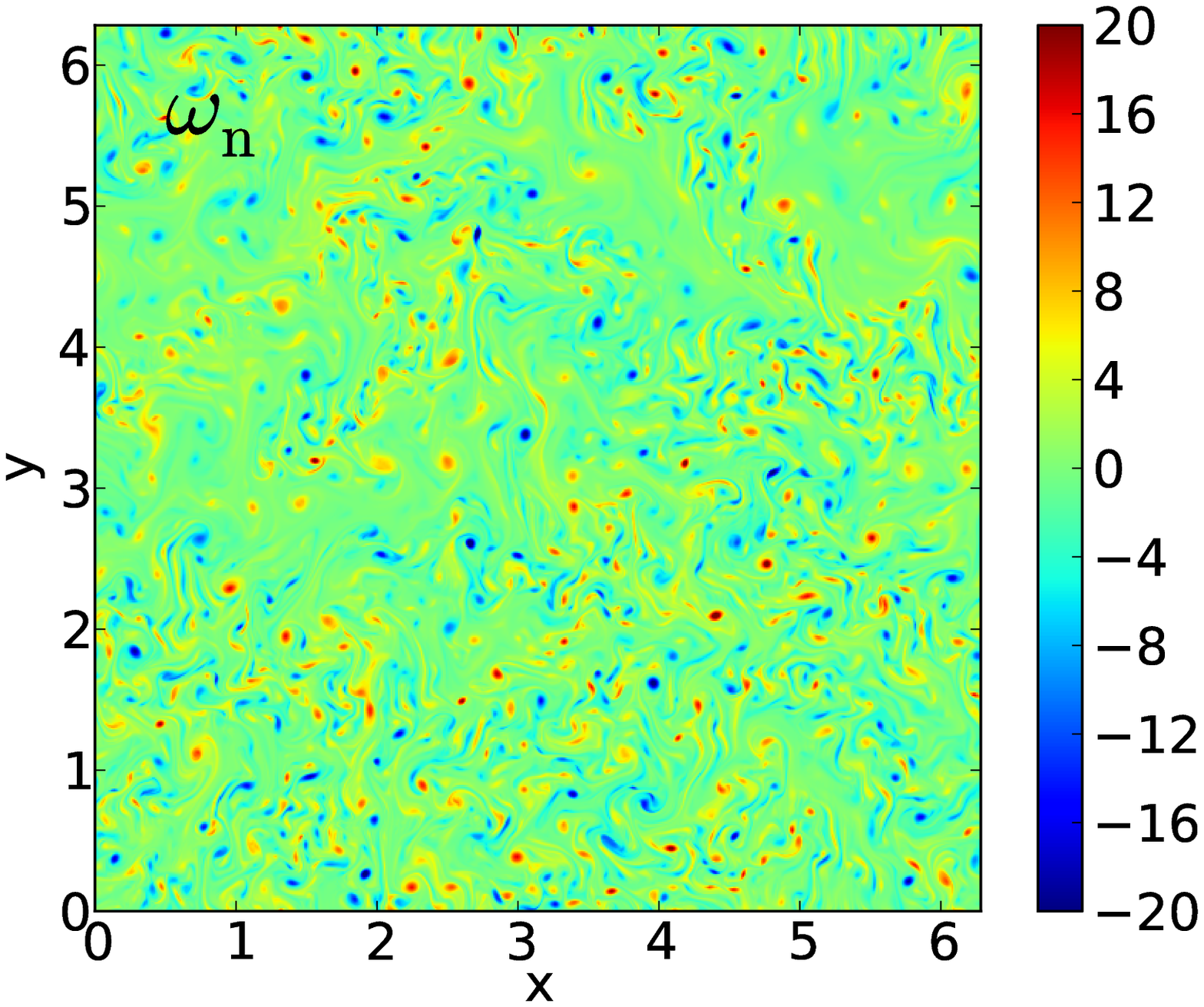}
\put(-75,170){\large{\bf (a)}}
\includegraphics[width=0.49\linewidth]{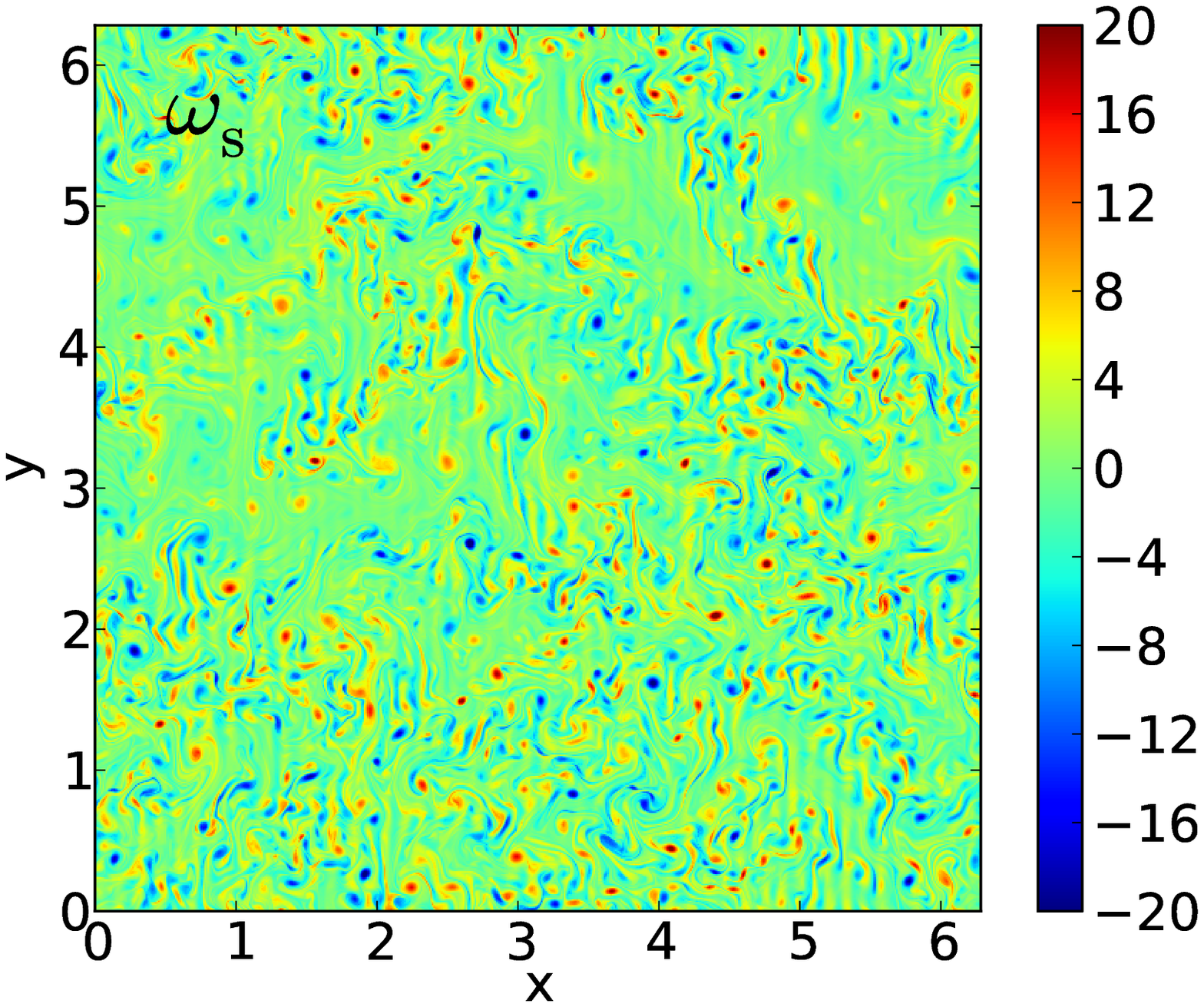}
\put(-75,170){\large{\bf (b)}}
}
\caption{(Color online) Pseudocolor plots of the vorticity fields, $\omega_n$
and $\omega_s$, from our DNS run $\tt R2a$ at $t=1500$ (panels (c) and (d), $k_{\rm
f}=50$); these plots show that the normal and superfluid
component are locked to each other.
}
\label{fig:snapvort}
\end{figure*}

In Fig.~\ref{fig:snapvort} we present the pseudocolor plots of 
$\omega_n$ and $\omega_s$ for run $\tt R2a$ (panels (a) and (b)). 

\begin{figure*}
\centering
\resizebox{0.5\linewidth}{!}{
\includegraphics[width=0.4\linewidth,unit=1mm]{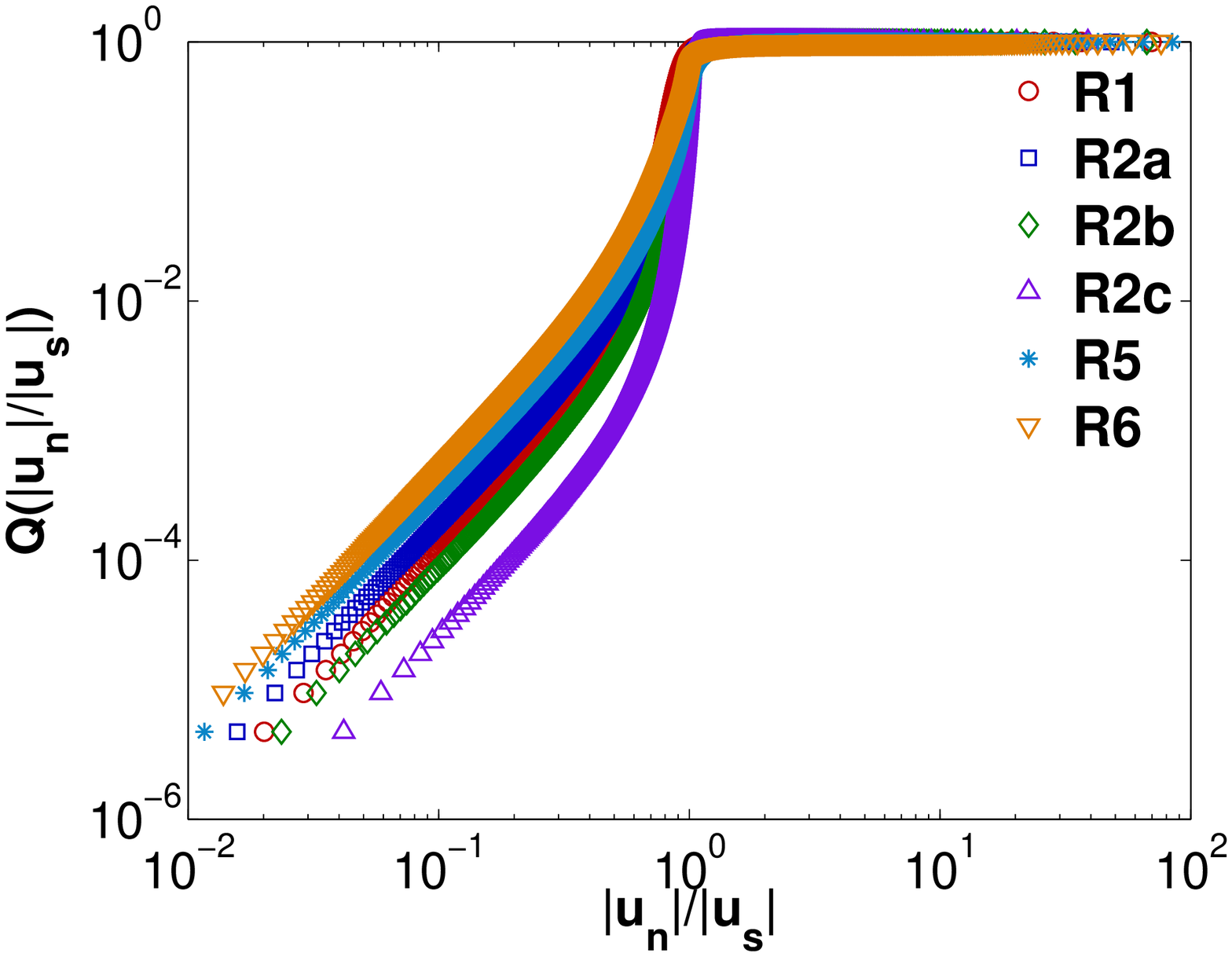}
}
\caption{(Color online) Log-log (base 10) plots of the cumulative distribution
functions (CDF) $Q(\gamma)$ of $\gamma=|\mathbf{u}_{\rm n}|/|\mathbf{u}_{\rm
s}|$ for the runs $\tt R1$,  $\tt R2a$-$\tt R2c$, $\tt R5$, and $\tt R6$. 
These CDFs show power-law tails
$Q(\gamma)\sim \gamma^2$, i.e., the PDF $P(\gamma) \sim \gamma$, for $\gamma \ll 1$.
}
\label{fig:pdfanglethetacmlratio}
\end{figure*}

In Fig.~\ref{fig:pdfanglethetacmlratio} we show
plots of the cumulative distribution functions (CDFs) $Q(\gamma)$ of
$\gamma=|\mathbf{u}_{\rm n}|/|\mathbf{u}_{\rm s}|$, 
for the runs $\tt R1$,  $\tt R2a$-$\tt R2c$, $\tt R5$, and $\tt R6$. 
These CDFs exhibit power-law tails that imply that the PDF
$P(\gamma) \sim \gamma$, for $\gamma \ll 1$; the power-law exponents of these tails of $P(\gamma)$ are
universal in the sense that they do not depend on $B$, $\rho_{\rm n}/\rho$, and
$k_{\rm f}$.

\begin{figure*}
\centering
\begin{overpic}
[width=0.25\linewidth,unit=1mm]{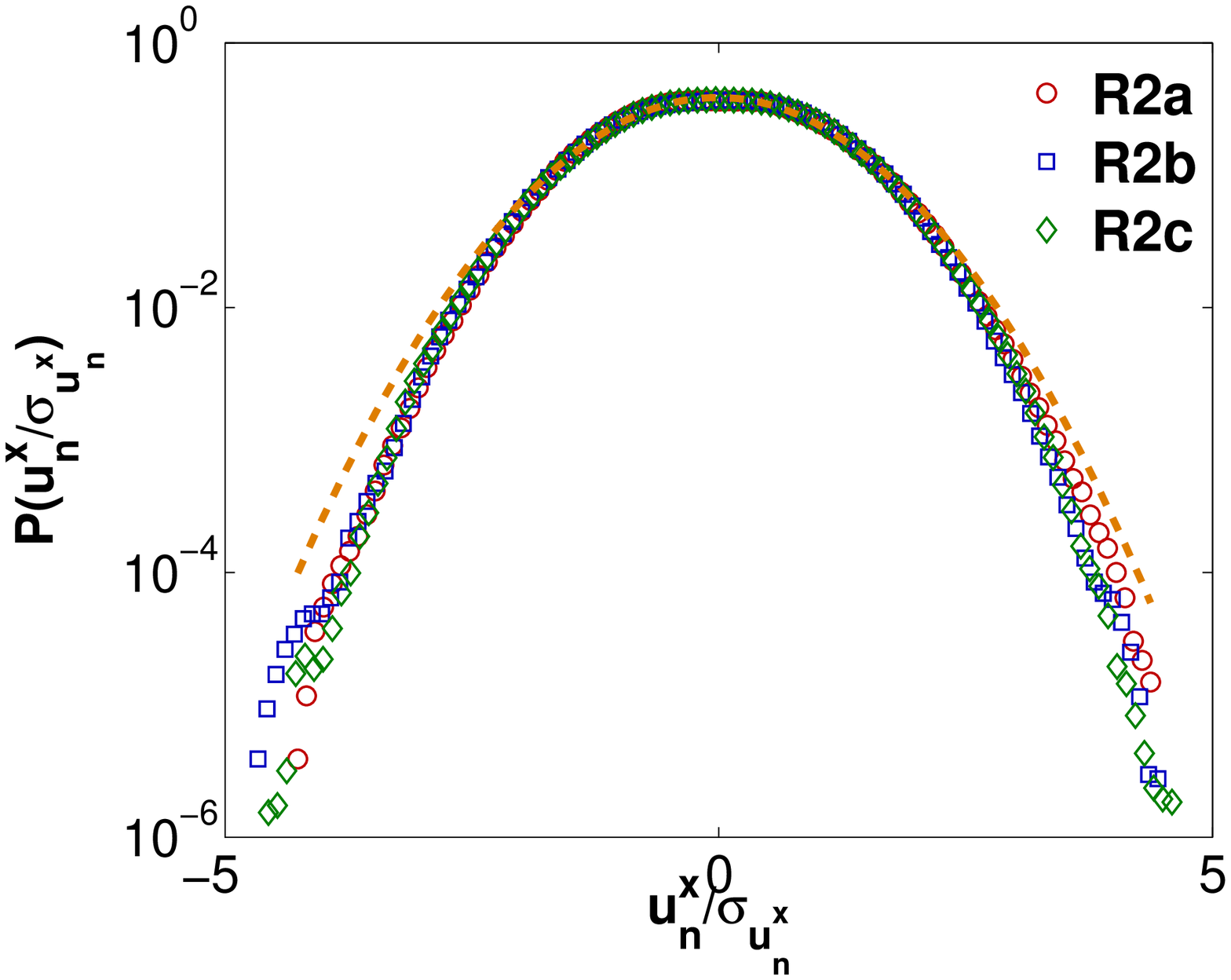}
\put(25,10){\large{\bf (a)}}
\end{overpic}
\begin{overpic}
[width=0.25\linewidth,unit=1mm]{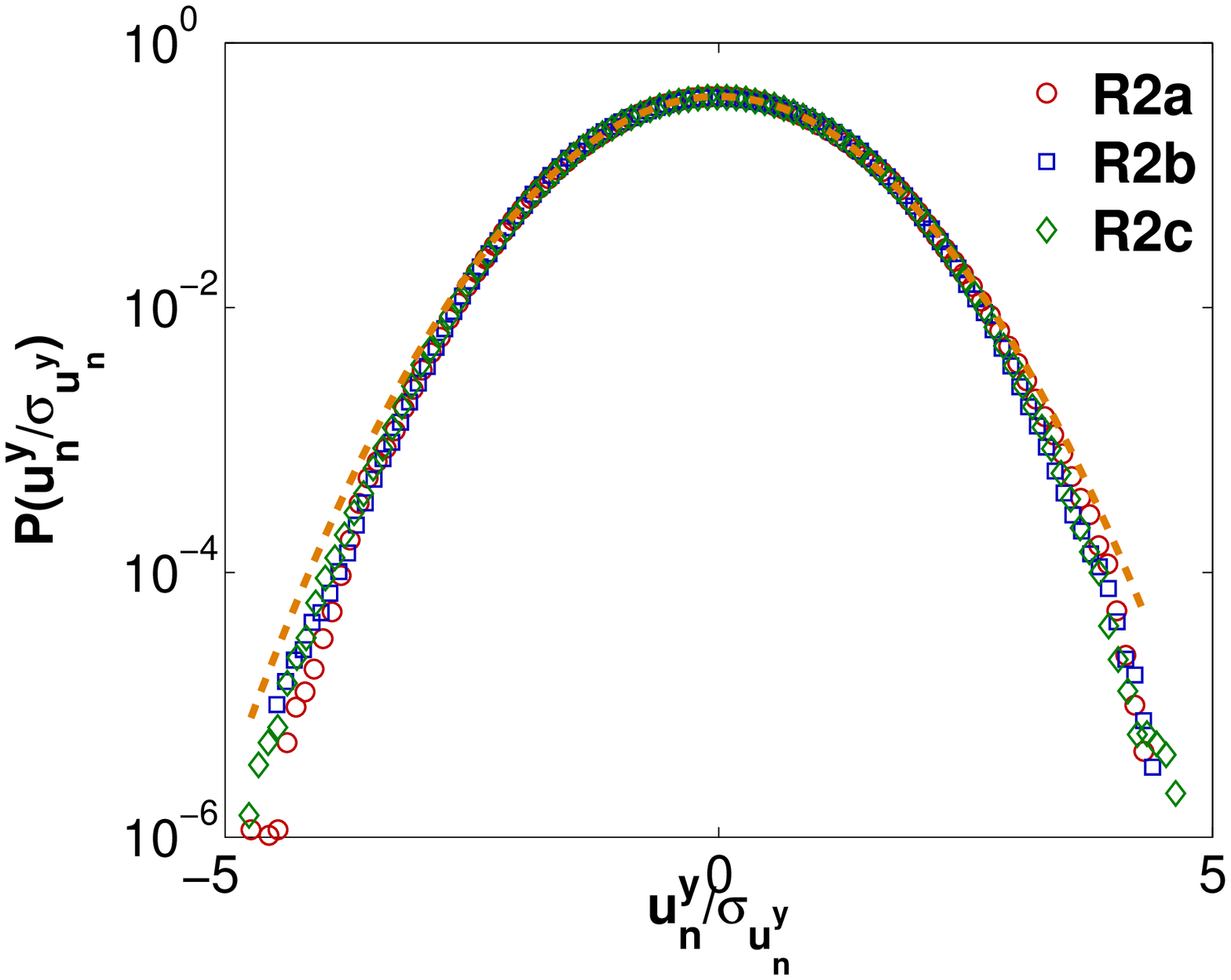} 
\put(25,10){\large{\bf (b)}}
\end{overpic}
\begin{overpic}
[width=0.25\linewidth,unit=1mm]{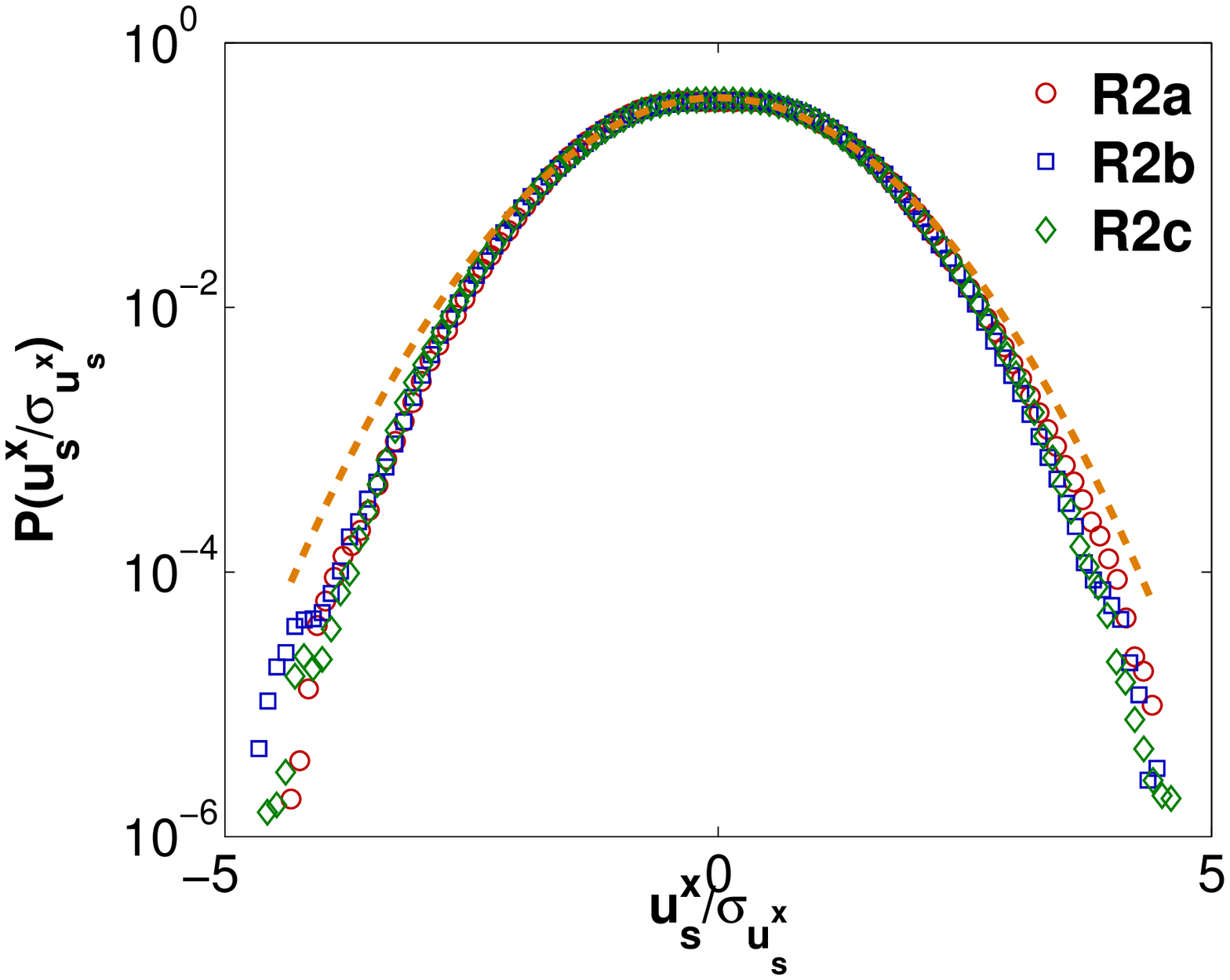}
\put(25,10){\large{\bf (c)}}
\end{overpic}
\begin{overpic}
[width=0.25\linewidth,unit=1mm]{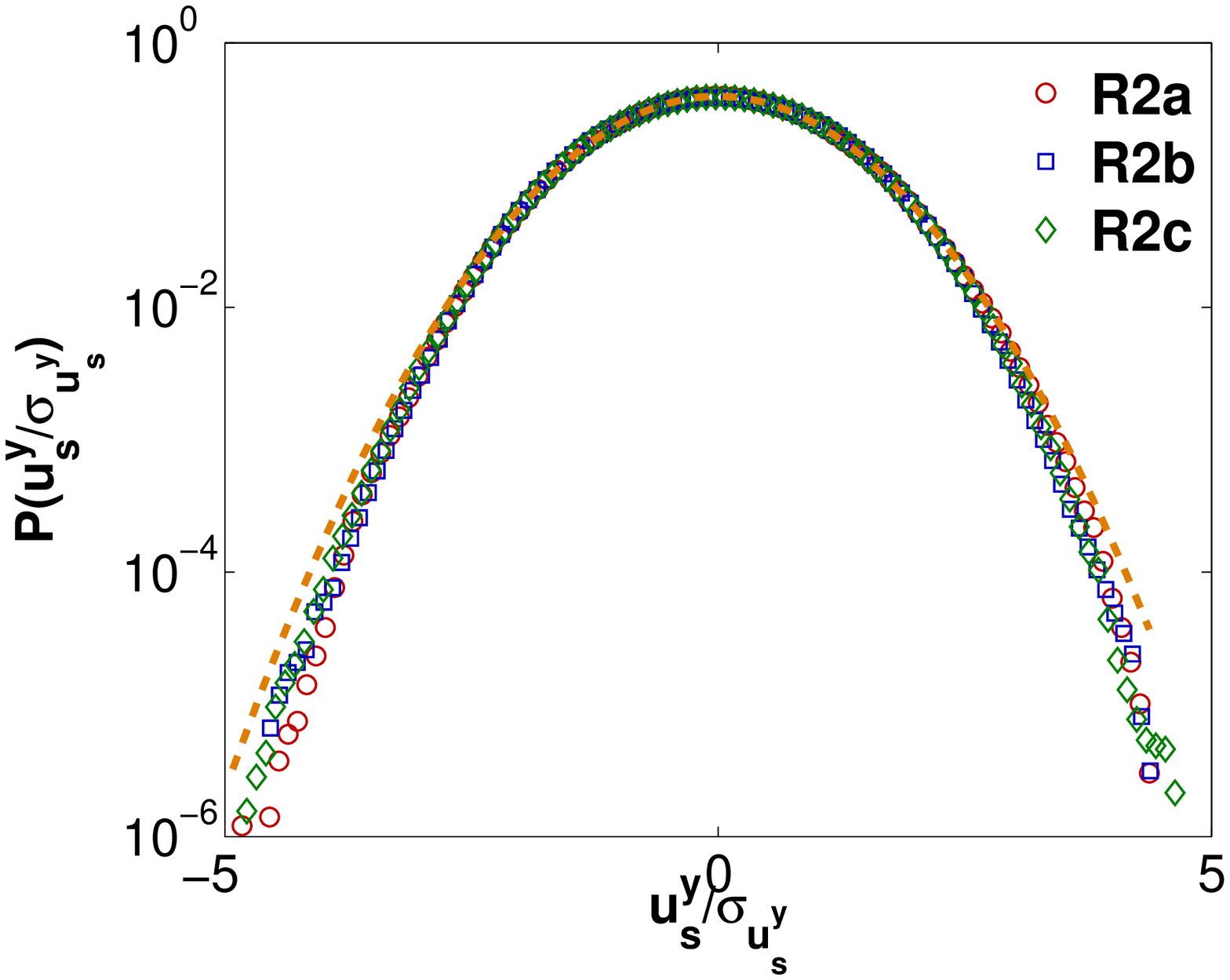}
\put(25,10){\large{\bf (d)}}
\end{overpic}
\caption{(Color online) Semilogarithmic (base 10) plots of the PDFs of the 
(a) $x$ component $\mathbf{u}_{\rm n}^{x}$ and 
(b) $y$ component $\mathbf{u}_{\rm n}^{y}$ of the normal fluid velocity;
PDFs of (c) $x$ component $\mathbf{u}_{\rm s}^{x}$;
(d) $y$ component $\mathbf{u}_{\rm s}^{y}$ of the superfluid velocity.
$\sigma_{\rm u^{\rm j}_{\rm i}}$ denotes the standard-deviation of the field 
$u^{\rm j}_{\rm i}$, here $i \in({\rm n,s})$ and $j \in({\rm x,y})$.
These data are from our DNS runs $\tt R2a$ (red circles), $\tt R2b$ (blue squares), and 
$\tt R2c$ (green diamonds), respectively; the orange dashed line indicates a Gaussian fit.
}
\label{fig:figpdfR2aR2bR2c}
\end{figure*}

\begin{figure*}
\centering
\begin{overpic}
[width=0.25\linewidth,unit=1mm]{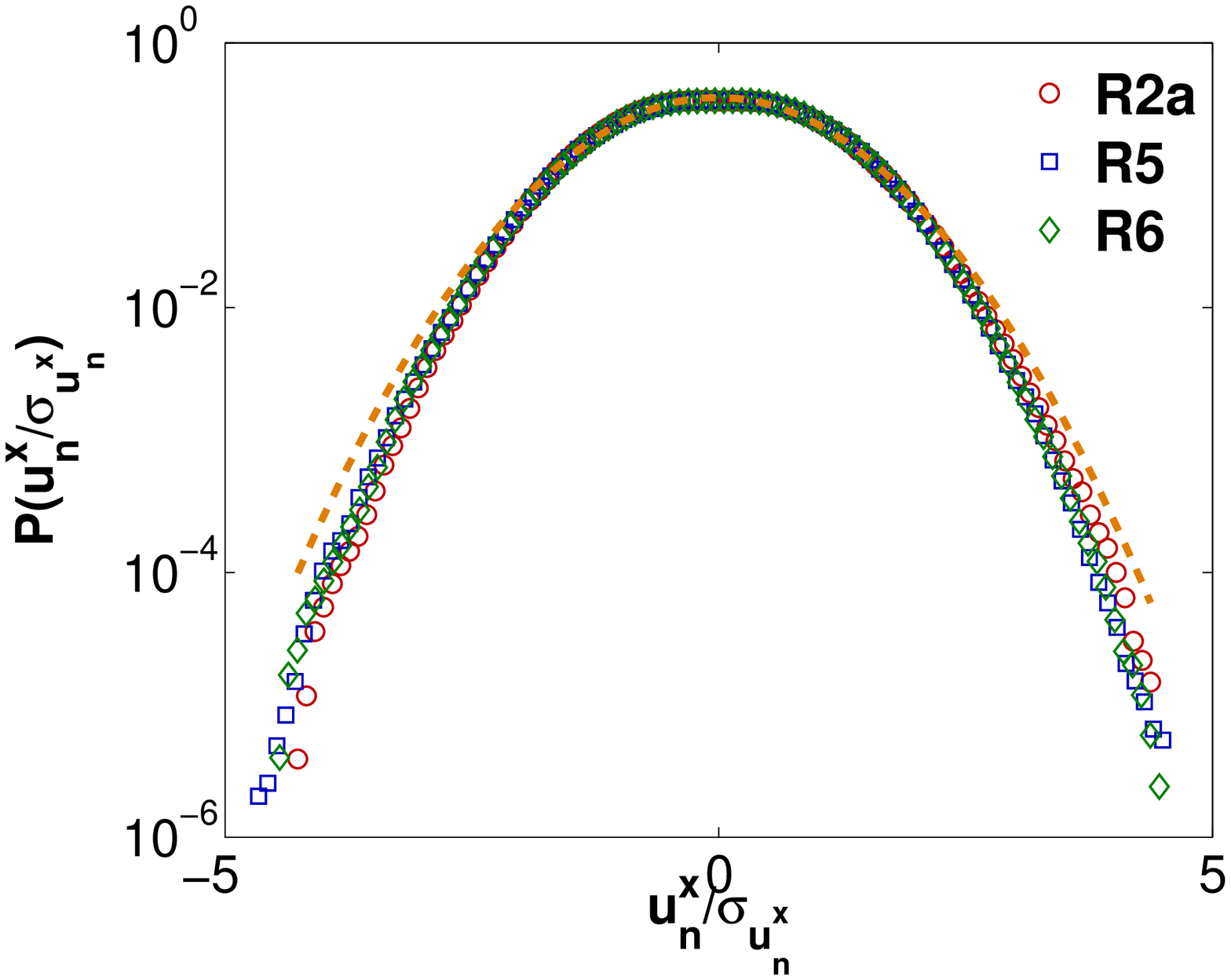}
\put(25,10){\large{\bf (a)}}
\end{overpic}
\begin{overpic}
[width=0.25\linewidth,unit=1mm]{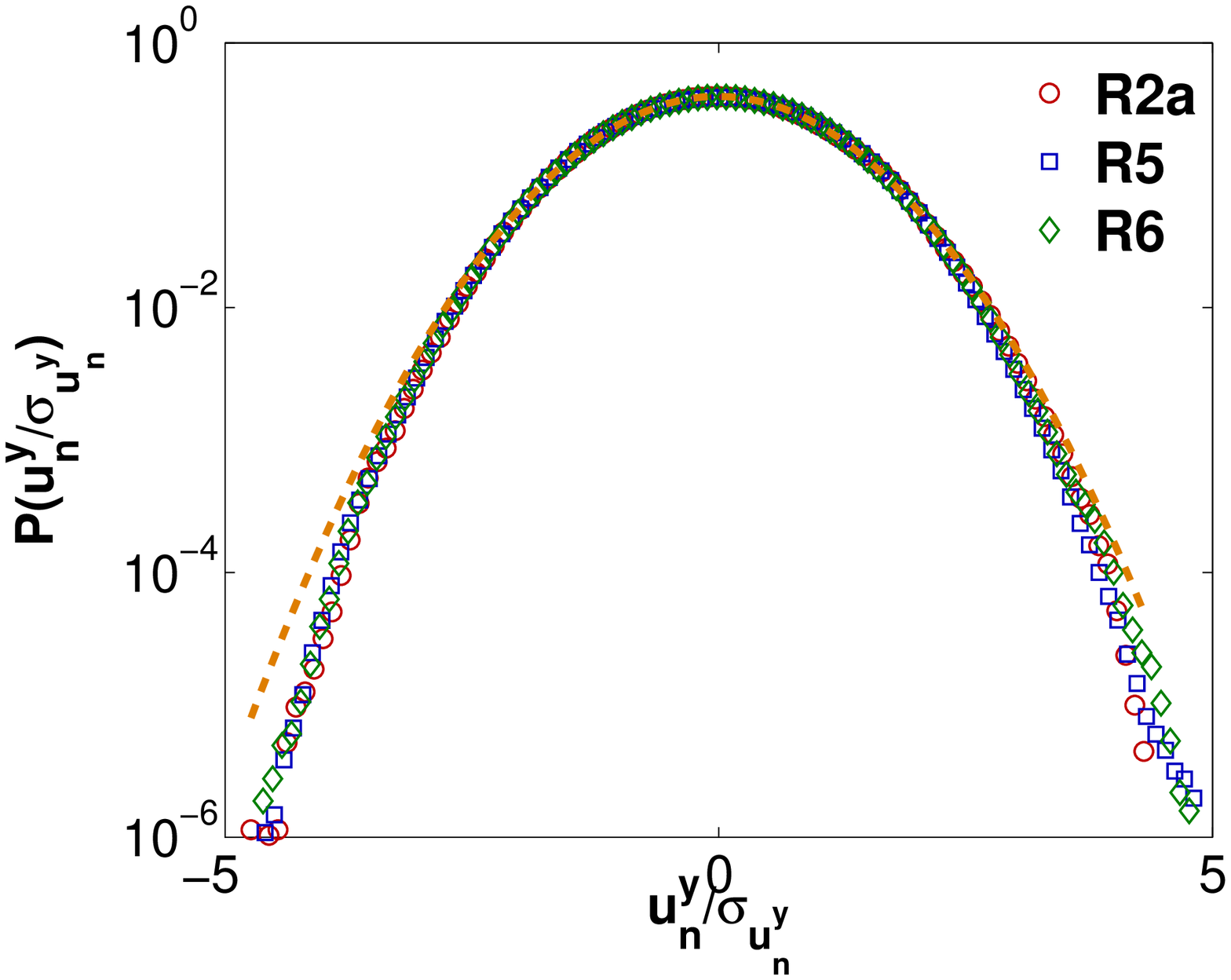}
\put(25,10){\large{\bf (b)}}
\end{overpic}
\begin{overpic}
[width=0.25\linewidth,unit=1mm]{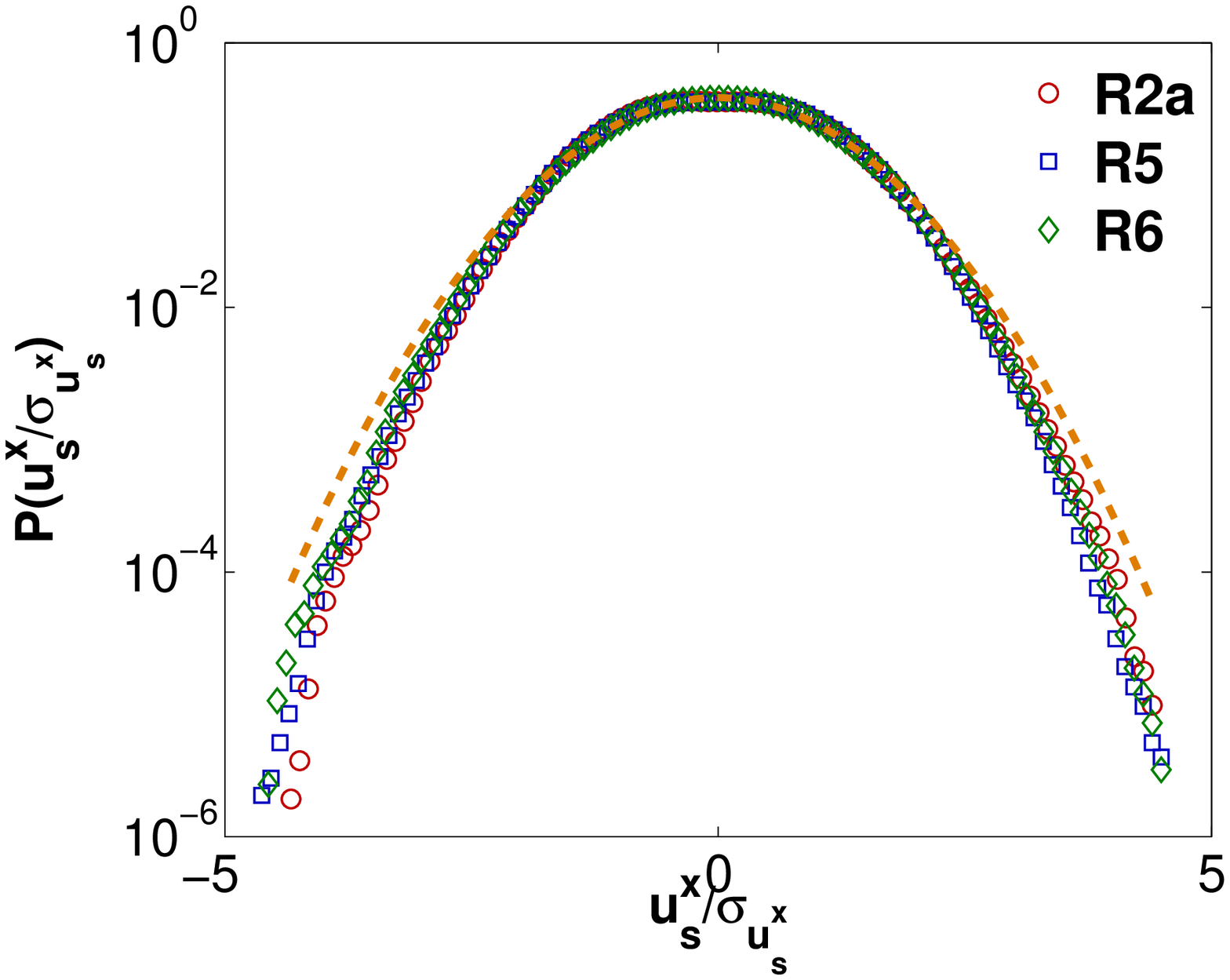}
\put(25,10){\large{\bf (c)}}
\end{overpic}
\begin{overpic}
[width=0.25\linewidth,unit=1mm]{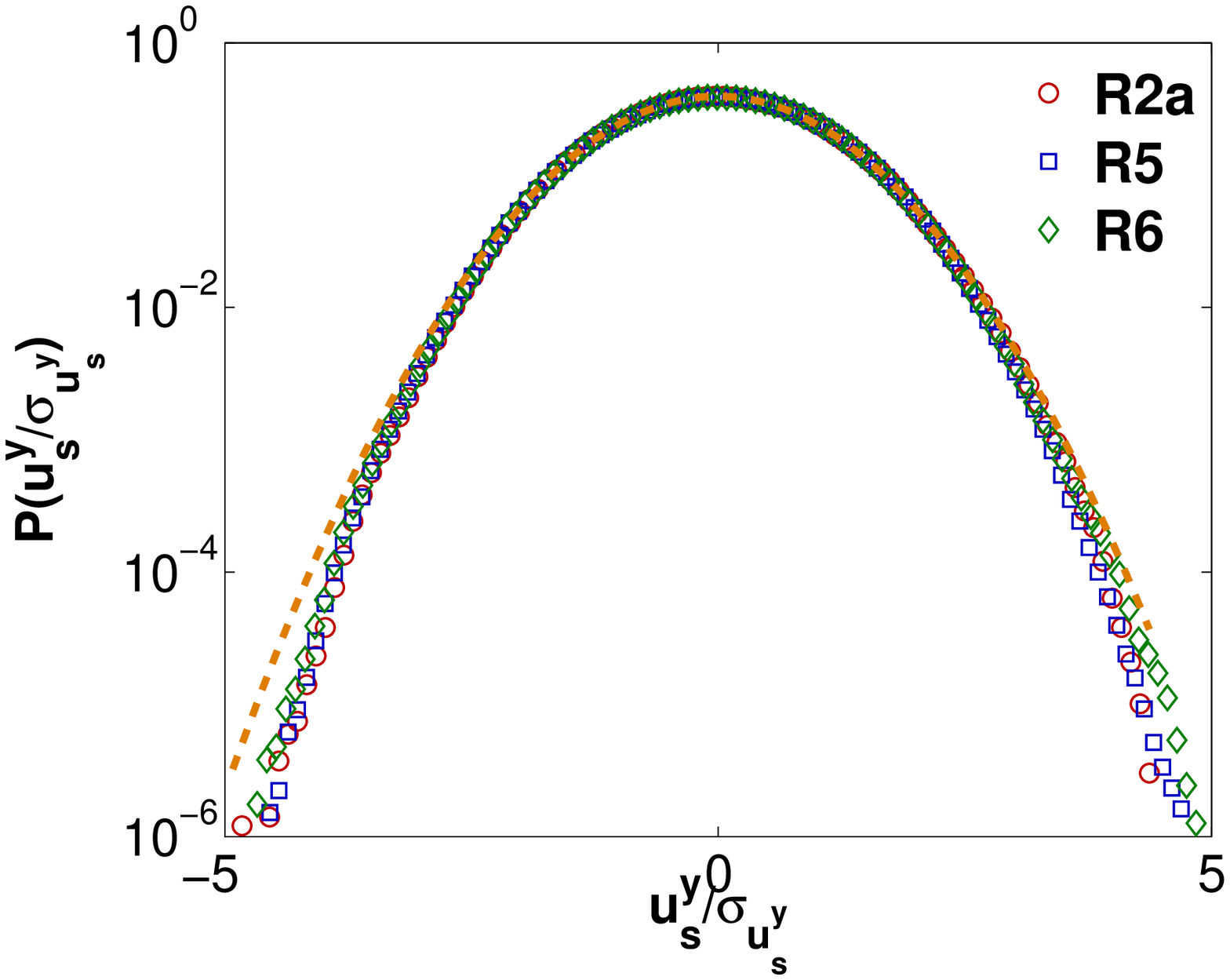}
\put(25,10){\large{\bf (d)}}
\end{overpic}
\caption{(Color online) Semilogarithmic (base 10) plots of the PDFs of the 
(a) $x$ component $\mathbf{u}_{\rm n}^{x}$ and 
(b) $y$ component $\mathbf{u}_{\rm n}^{y}$ of the normal fluid velocity;
PDFs of (c) $x$ component $\mathbf{u}_{\rm s}^{x}$;
(d) $y$ component $\mathbf{u}_{\rm s}^{y}$ of the superfluid velocity.
$\sigma_{\rm u^{\rm j}_{\rm i}}$ denotes the standard-deviation of the field 
$u^{\rm j}_{\rm i}$, here $i \in({\rm n,s})$ and $j \in({\rm x,y})$.
These data are from our DNS runs $\tt R2a$ (red circles), $\tt R5$ (blue squares), and 
$\tt R6$ (green diamonds), respectively; the orange dashed line indicates a Gaussian
fit.}
\label{fig:pdfvelR2aR5R6}
\end{figure*}

\begin{figure*}
\centering
\begin{overpic}
[width=0.25\linewidth,unit=1mm]{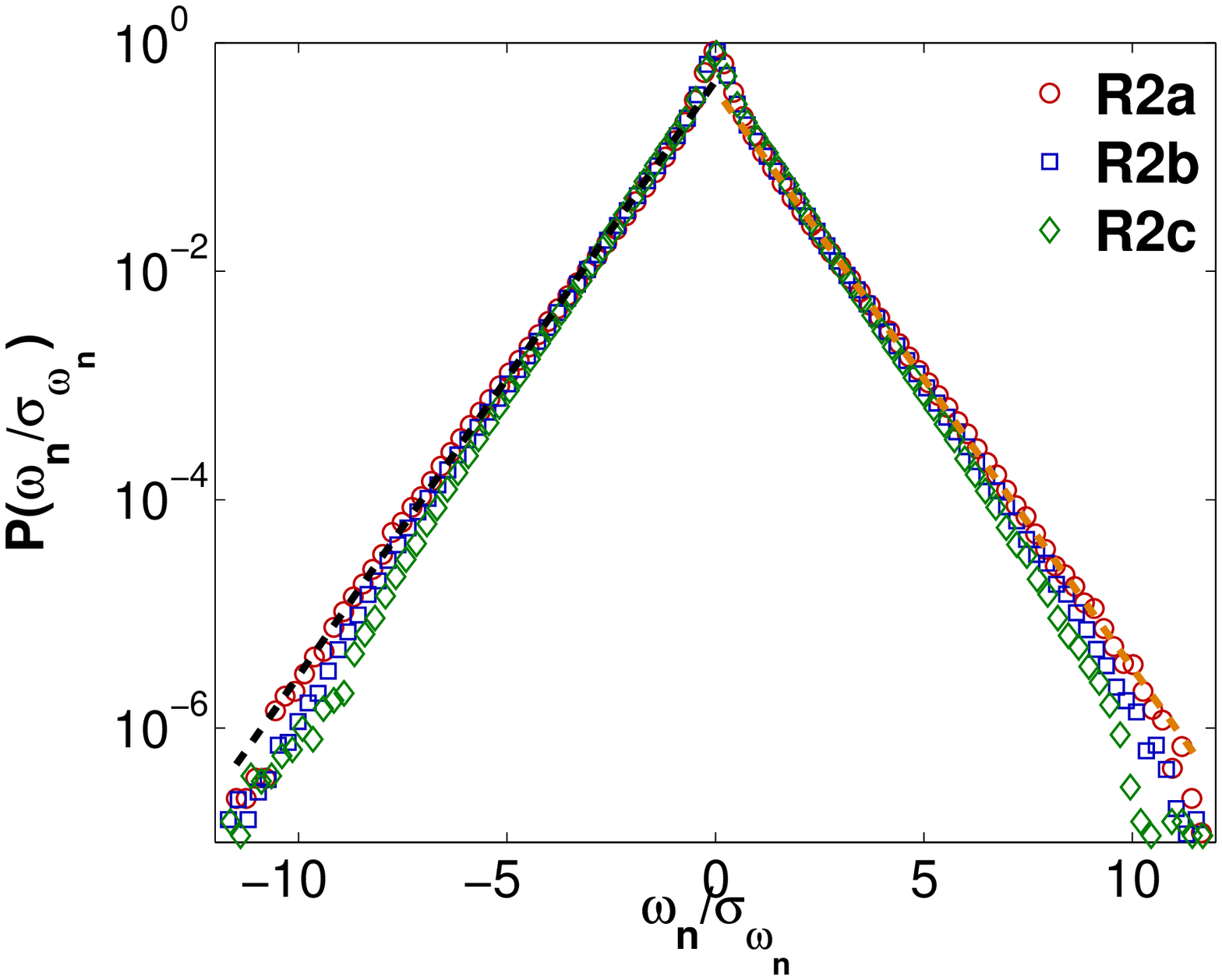}
\put(25,10){\large{\bf (a)}}
\end{overpic}
\begin{overpic}
[width=0.25\linewidth,unit=1mm]{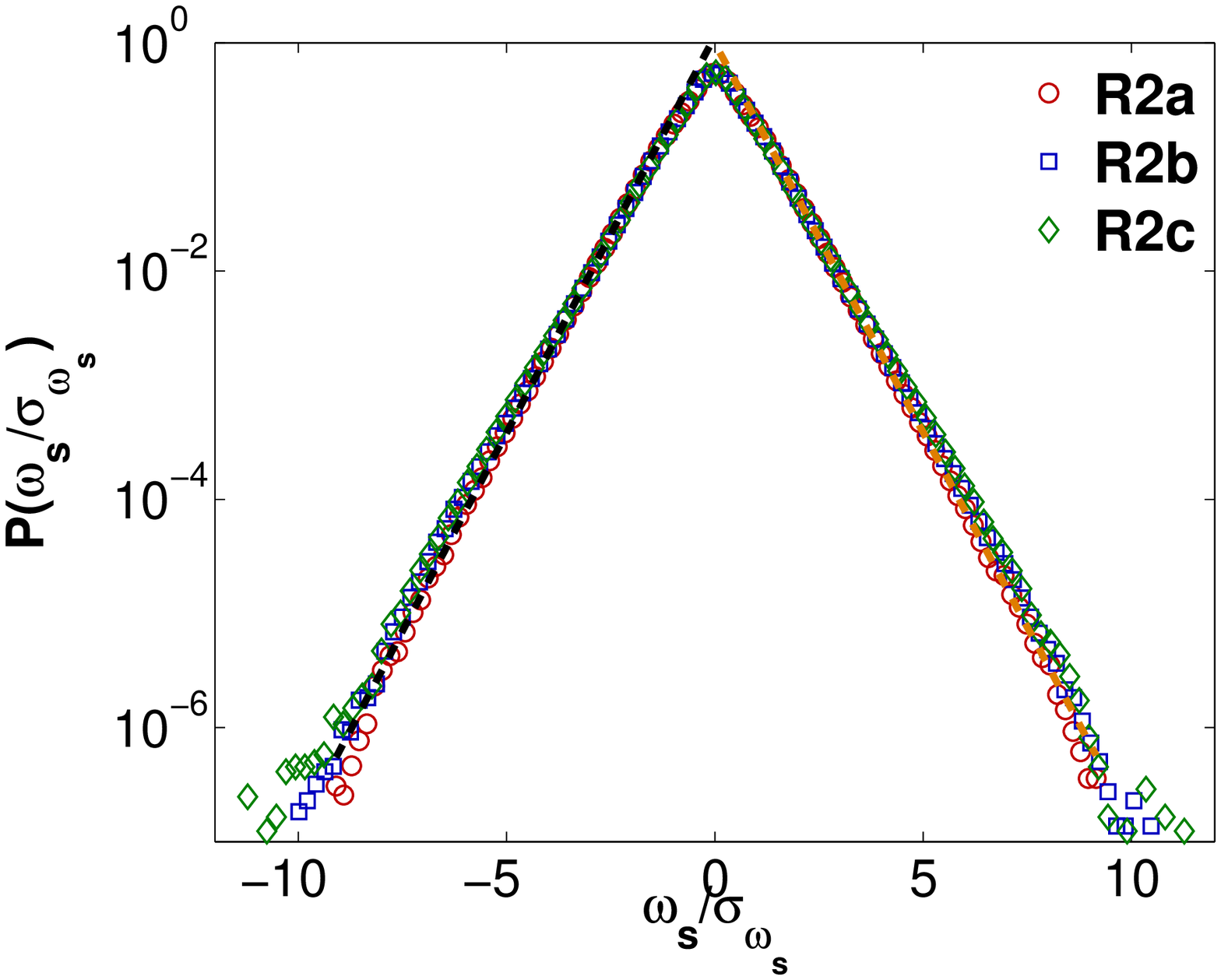}
\put(25,10){\large{\bf (b)}}
\end{overpic}
\begin{overpic}
[width=0.25\linewidth,unit=1mm]{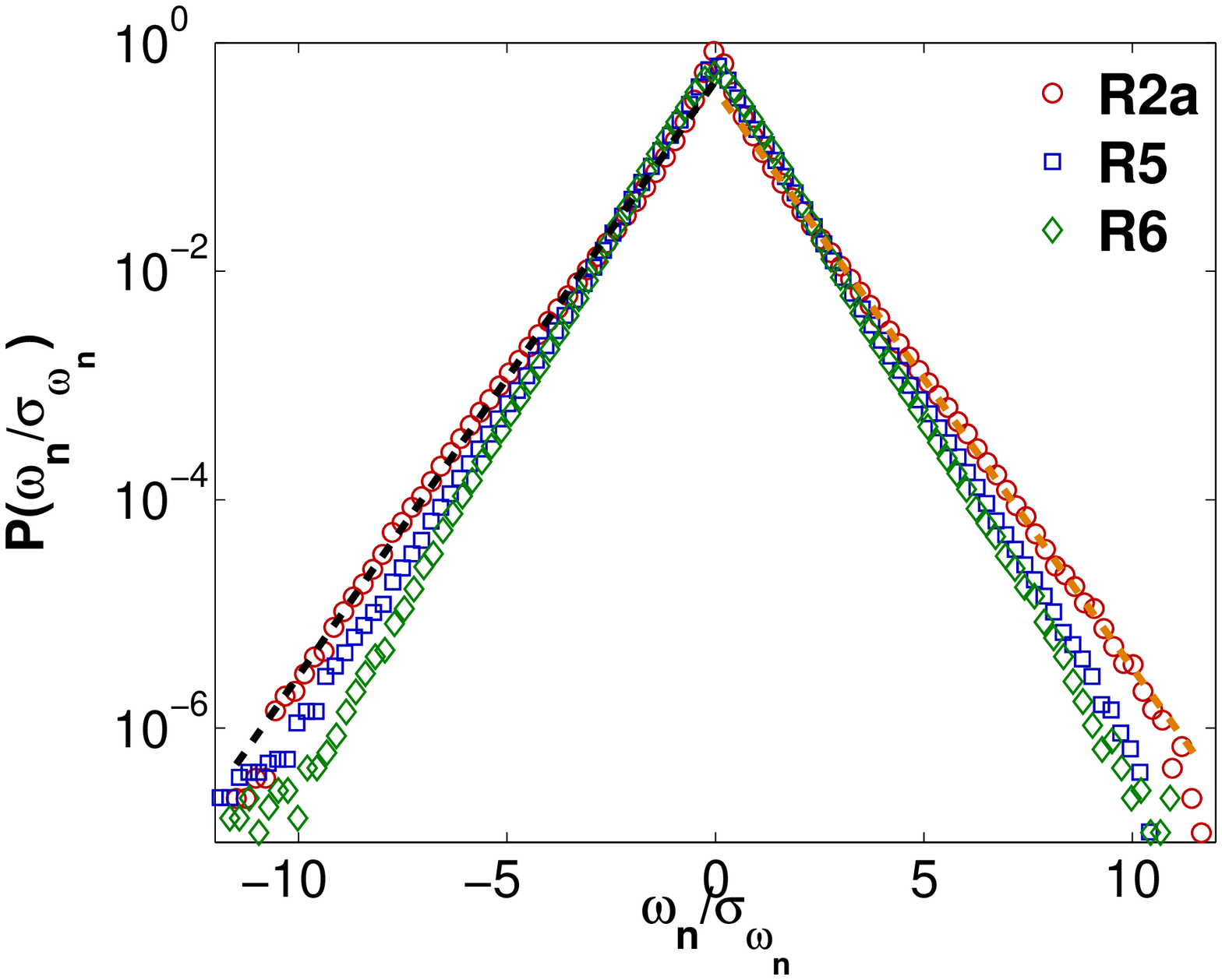}
\put(25,10){\large{\bf (c)}}
\end{overpic}
\begin{overpic}
[width=0.25\linewidth,unit=1mm]{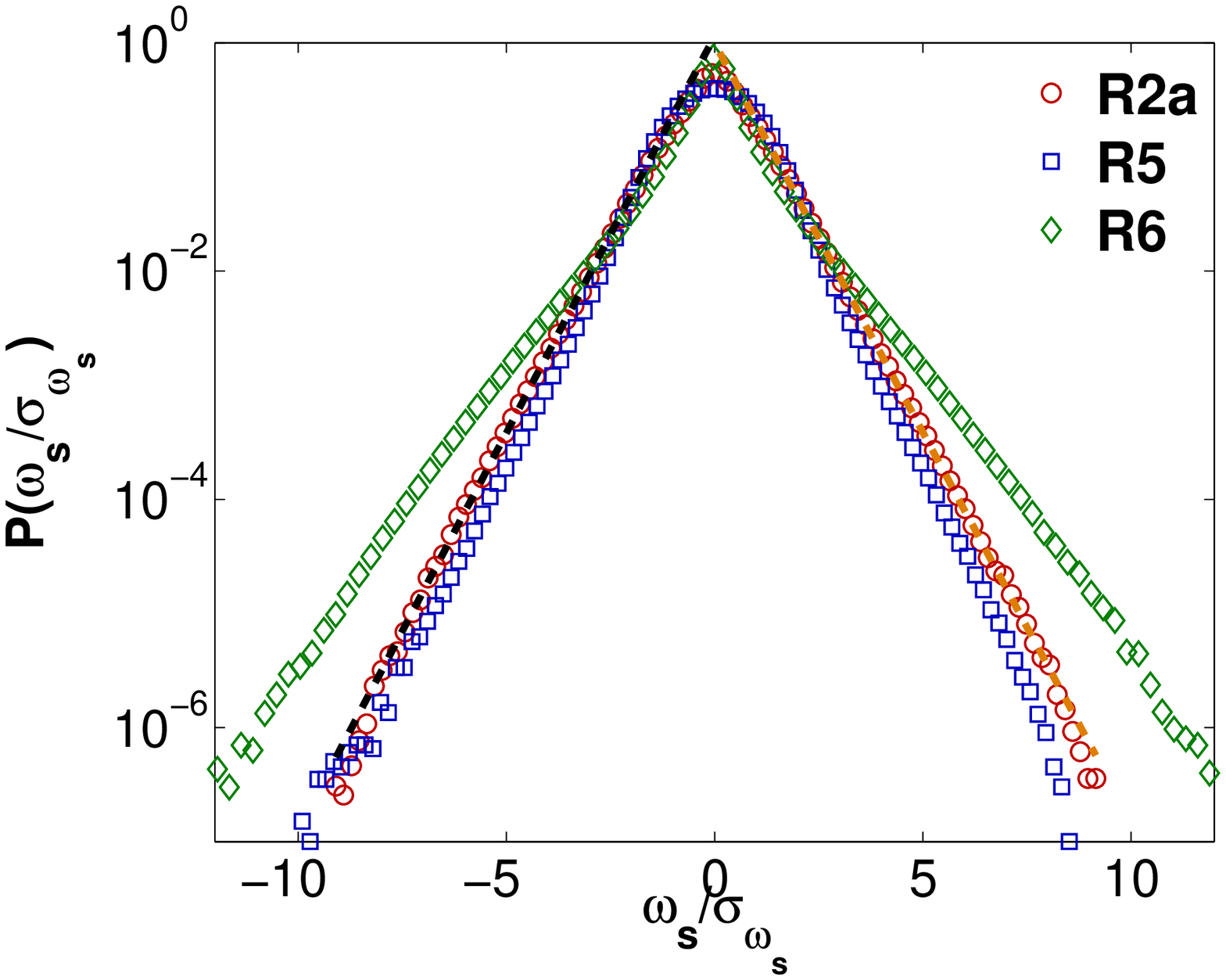}
\put(25,10){\large{\bf (d)}}
\end{overpic}
\caption{(Color online) Semilogarithmic (base 10) plots of the PDFs of the vorticity of the 
(a) normal fluid
($\omega_{\rm n}$) from our DNS runs $\tt R2a$ (red circles), $\tt R2b$ (blue squares), 
and $\tt R2c$ (green diamonds); the black- and the orange-dashed lines indicate an exponential fit to
the left (slope $=-0.5064$) and the right (slope $=0.5207$) branches of the PDF 
$P(\omega_n/\sigma_{\omega_{n}})$ for the DNS run $\tt R2a$; 
PDFs of the (b) superfluid ($\omega_{\rm s}$) from our DNS runs $\tt R2a$ (red circles), 
$\tt R2b$ (blue squares), and $\tt R2c$ (green diamonds); the black- and the orange-dashed 
lines indicate an exponential fit to the left (slope $=-0.6818$) and the right 
(slope $=0.6951$) branches of the PDF $P(\omega_s/\sigma_{\omega_{s}})$ for the DNS run $\tt R2a$;
PDFs of the (c) normal fluid
($\omega_{\rm n}$) from our DNS runs $\tt R2a$ (red circles), $\tt R5$ (blue squares), 
and $\tt R6$ (green diamonds); the black- and the orange-dashed lines indicate an exponential fit to
the left (slope $=-0.5064$) and the right (slope $=0.5207$) branches of the PDF 
$P(\omega_n/\sigma_{\omega_{n}})$ for the DNS run $\tt R2a$; 
PDFs of the (d) superfluid ($\omega_{\rm s}$) from our DNS runs $\tt R2a$ (red circles), 
$\tt R5$ (blue squares), and $\tt R6$ (green diamonds); the black- and the orange-dashed 
lines indicate an exponential fit to the left (slope $=-0.6818$) and the right 
(slope $=0.6951$) branches of the PDF $P(\omega_s/\sigma_{\omega_{s}})$ for the DNS run $\tt R2a$.
$\sigma_{\omega_{\rm i}}$ denotes the standard-deviation of the field $\omega_{\rm i}$, 
here $i \in({\rm n,s})$.
}
\label{fig:figpdfomg}
\end{figure*}

\begin{figure*}
\centering
\includegraphics[width=0.48\linewidth]{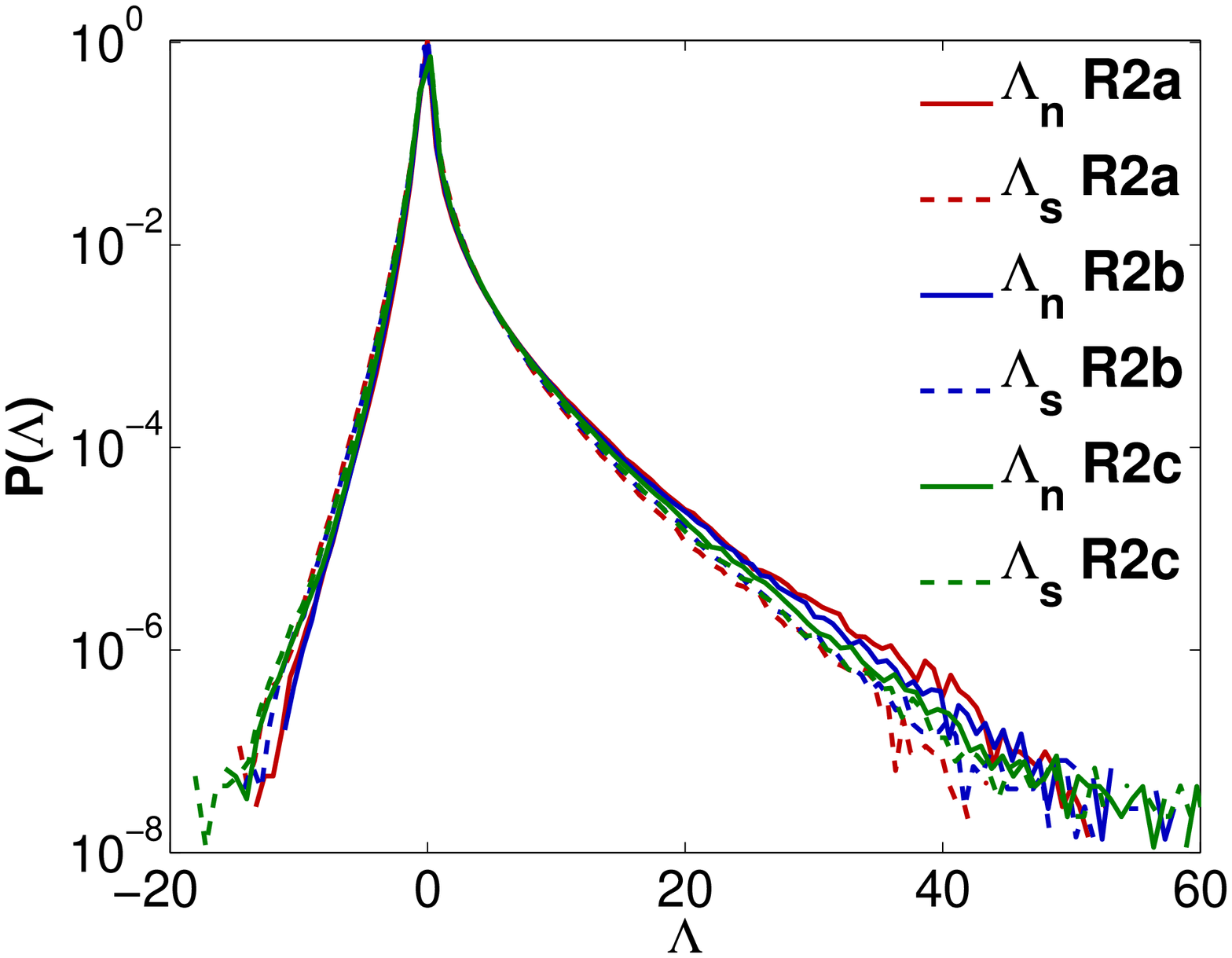}
\put(-100,35){\large{\bf (a)}}
\includegraphics[width=0.48\linewidth]{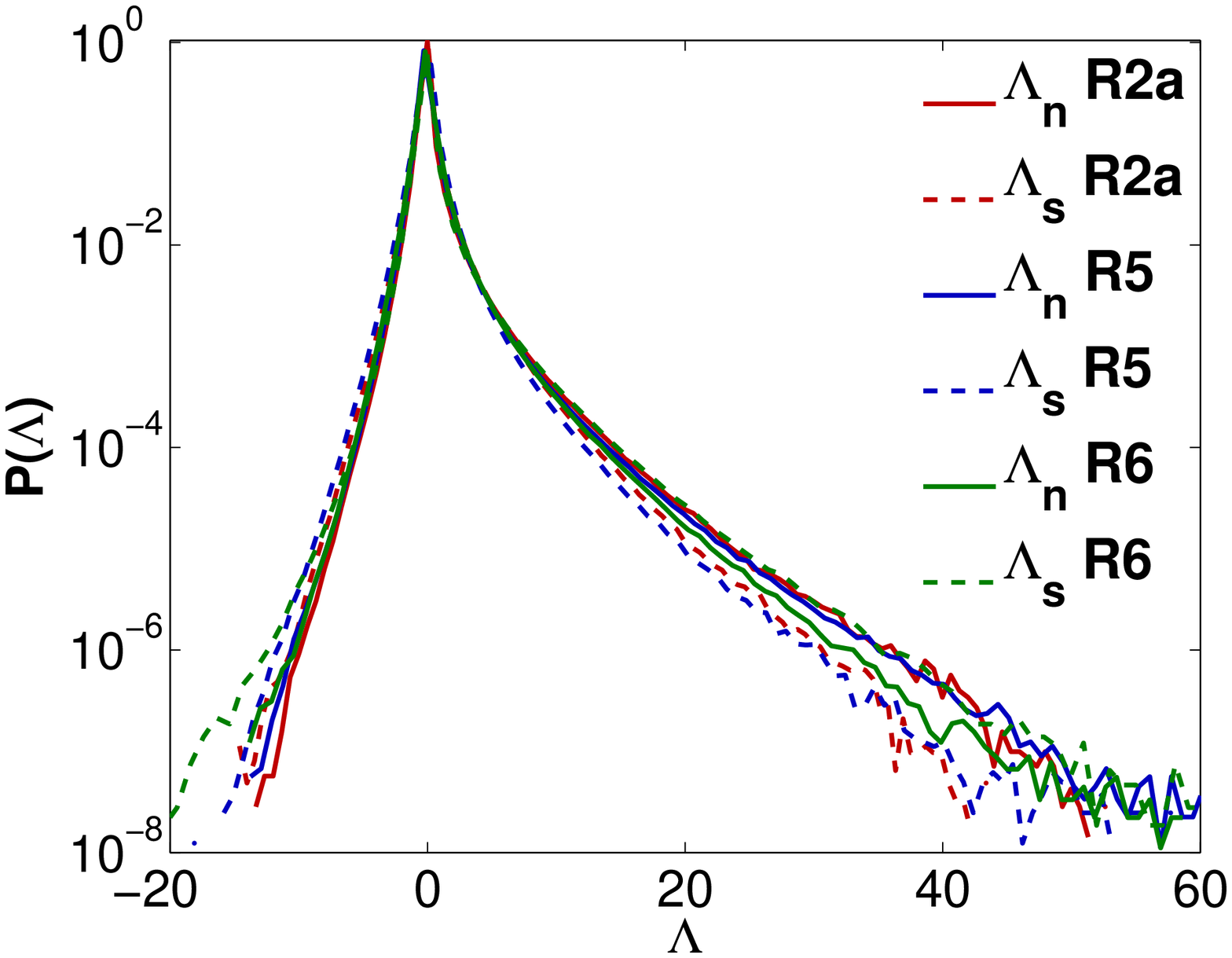}
\put(-100,35){\large{\bf (b)}}
\caption{(Color online) Semilogarthimic (base 10) plots of the PDFs of the Okubo-Weiss parameter for the normal
fluid $\Lambda_{\rm n}$ (solid line) and the superfluid $\Lambda_{\rm s}$ (dashed line);
(a) $\tt R2a$ (red line), $\tt R2b$ (blue line), and $\tt R2c$ (green line) showing the 
variation of these PDF with the mutual-friction coefficients $B=1$, $B=2$, and $B=5$, 
respectively; (b) $\tt R2a$ (red line), $\tt R5$ (blue line), and $\tt R6$ (green line) 
showing the variation of these PDF with the $\rho_{\rm n}/\rho = 0.1$, $\rho_{\rm n}/\rho = 0.5$, 
and $\rho_{\rm n}/\rho = 0.9$, respectively.
In 2D, classical-fluid turbulence, the PDF of the Okubo-Weiss parameter is qualitatively
similar~\cite{perlekarnjp2dfilms} to the PDFs shown here.
}
\label{fig:pdflambda}
\end{figure*}

In Figs.~\ref{fig:figpdfR2aR2bR2c} and~\ref{fig:pdfvelR2aR5R6} we show that 
the PDFs of the Cartesian components of the normal and superfluid velocities 
in 2D HVBK turbulence are close to Gaussian (as in 2D, classical-fluid turbulence).
Figure~\ref{fig:figpdfomg} shows that the tails of the PDFs of the normal and superfluid vorticity in
2D HVBK turbulence are close to exponentials, as in 2D, classical-fluid turbulence.
In Fig.~\ref{fig:pdflambda} we show the PDFs of the Okubo-Weiss parameter
$\Lambda_i=(\omega_i^2-\sigma_i^2)/4$, $i\in(n,s)$,
whose sign determines whether the flow in a given region
is vortical ($\Lambda_i>0$) or strain-dominated ($\Lambda_i<0$); $\omega_i^2$ and $\sigma_i^2$ 
are the squares of the vorticity and the strain-rate, respectively. 
These PDFs are similar to their 2D, classical-fluid-turbulence counterparts.
 
\end{document}